\documentclass[fleqn]{article}
\usepackage{mystyle,mymath,proof,latexsym,amssymb}
\usepackage{color}
\usepackage[usenames,dvipsnames]{xcolor}

\begin{document}
\sloppy \hbadness=10000 \vbadness=10000


\def\calP{{\cal P}}
\def\Init{{\rm init}}
\def\Bounded{{\rm Bounded}}


\def\BFactor{{\rm Bounded Factor}}
\def\Ncell{{n_{\rm cell}}}
\def\Out{{\rm Out}}
\def\Amalgamation{{\rm Amalg}}
\def\toto{\twoheadrightarrow}
\def\toeq{\Rightarrow}
\def\Cone{{\rm cone}}
\def\Subcone{{\rm subcone}}
\def\Un{{\rm un}}
\def\Tr{{\rm tr}}
\def\Bk{{\rm bk}}
\def\Dep{{\rm Dep}}
\def\DefC{{\rm DefC}}
\def\|{$|$}


\newdimen\Programindent
\Programindent=3ex

\def\Hrule{\vskip 3pt\hrule\vskip 3pt}
\long\def\FunctionL#1#2#3{{\parindent=0pt{\bf function} {\rm #1}$(#2)$
{{\advance\parindent by \Programindent
#3
}
\par
{\bf end function}
}}}

\long\def\LoopL#1{{\bf loop}
{{\advance\parindent by \Programindent
#1
}
\par
{\bf end loop}
}}

\long\def\WhileL#1#2{{\bf while} $#1$ {\bf do}
{{\advance\parindent by \Programindent
#2
}
\par
{\bf end while}
}}

\long\def\ForeachL#1#2#3{{\bf for each} $#1$ {\bf in} $#2$ {\bf do}
{{\advance\parindent by \Programindent
#3
}
\par
{\bf end for}
}}

\long\def\ForeachS#1#2#3{{\bf for each} $#1$ {\bf in} $#2$ {\bf do}
{
#3
}
}

\def\IfthenS#1#2{{\bf if} {#1} {\bf then} {#2}\par}
\long\def\IfthenL#1#2{{\bf if} {#1} {\bf then}\par
{\advance\parindent by \Programindent
#2
}
\par
{\bf end if}
}

\def\ElseS#1{{\bf else} {#1}\par}
\long\def\ElseL#1{{\bf else}\par
{\advance\parindent by \Programindent
#1
}
\par
{\bf end if}
}

\def\ElseifthenS#1#2{{\bf else if} {#1} {\bf then} {#2}\par}
\long\def\ElseifthenL#1#2{{\bf else if} {#1} {\bf then}\par
{\advance\parindent by \Programindent
#2
}
\par
}

\def\WhereS#1{{\bf where} {#1}}
\def\LetS#1#2{{\bf let} $#1$ {\bf be} $#2$}
\long\def\LetL#1#2{{\bf let} $#1$ {\bf be} $#2$}
\def\BReak{{\bf break}}
\def\COntinue{{\bf continue}}
\def\FAil{{\bf fail}}
\def\SUccess{{\bf success}}
\def\Input{{\bf input:\ }}
\def\Return#1{{\bf return} $#1$}
\def\GOto#1{{\bf goto} {\rm #1}}


\def\LsO{{{\rm lsO}}}
\def\LsE{{{\rm lsE}}}


\def\Swand{{\Wand^s}}
\def\Identity{{{\rm identity}}}
\def\CSH{{\rm CSH}}
\def\Widen{{\rm widen}}
\def\Factor{{\rm Factor}}
\def\Dll{{\rm dll}}
\def\Listnest{{\rm listnest}}
\def\Lsnest{{\rm lsnest}}
\def\Skl#1{{\rm skl{#1}}}
\def\CV{{{\rm CV}}}
\def\IV{{{\rm IV}}}
\def\Dw{{\rm d_{wand}}}
\def\WAnd{{\rm wand}}


\def\Over#1#2{\deduce{#2}{#1}}
\def\List{{\rm List}}
\def\ListX{{\rm ListX}}
\def\ListO{{\rm ListO}}
\def\ListE{{\rm ListE}}
\def\Nl{{\rm nl}}


\def\SLLID{{\rm SLID}}
\def\SLID{{\rm SLID}}
\def\CSLID{{\rm CSLID}}
\def\CSLIDomega{{{\rm CSLID}^\omega}}
\def\Eclass{{\rm Eclass}}
\def\Eq{{\rm Eq}}
\def\Deq{{\rm Deq}}
\def\Satom{{\rm Satom}}
\def\Cells{{\rm Cells}}
\def\Roots{{\rm Roots}}
\def\Elim{{\rm Elim}}
\def\Case{{\rm Case}}
\def\Unfold{{\rm Unfold}}
\def\Pred{{\rm Pred}}
\def\Start{{\rm Start}}
\def\Unsat{{\rm Unsat}}
\def\EmpUnsat{{\rm empUnsat}}
\def\Split{{\rm Split}}
\def\Underscore{\underline{\phantom{x}}}
\def\Rootcell{{\rm rootcell}}
\def\Rootshape{{\rm rootshape}}
\def\Jointleaf{{\rm jointleaf}}
\def\Jointleafimplicit{{\rm jointleafimplicit}}
\def\Jointnode{{\rm jointnode}}
\def\Directjoint{{\rm directjoint}}


\def\Range{{\rm Range}}
\def\Xstart{X_{{\rm start}}}
\def\kstart{k_{{\rm start}}}
\def\Cstart{C_{{\rm start}}}
\def\Leaves{{\rm Leaves}}
\def\PureDist{{\rm PureDist}}
\def\Pure{{\rm Pure}}
\def\Identity{{\rm Identity}}
\def\Subst{{\rm Subst}}


\def\Connected{{\rm Connected}}
\def\Eststablished{{\rm Eststablished}}
\def\Valued{{\rm Valued}}


\def\SLRDbtw{{\rm SLRD}_{btw}}
\def\BTW{{{\rm BTW}}}
\def\Loc{{{\rm Loc}}}
\def\Ne{{\ne}}
\def\Nil{{{\rm nil}}}
\def\eqDef{=_{{\rm def}}}
\def\Inf#1{{\infty_{#1}}}
\def\Stores{{\rm Stores}}
\def\SVars{{{\rm SVars}}}
\def\Val{{{\rm Val}}}
\def\MSO{{{\rm MSO}}}
\def\Sep{{{\rm Sep}}}
\def\Septwo{{{\rm SLMI}}}
\def\SLMI{{{\rm SLMI}}}
\def\Sepinf{{\rm Sep}\infty}
\def\THeaps{{\rm THeaps}}
\def\Root{{\rm Root}}
\def\TG{{\rm TGraph}}


\def\Nil{{{\rm nil}}}
\def\eqDef{=_{{\rm def}}}
\def\Inf#1{{\infty_{#1}}}
\def\Stores{{\rm Stores}}
\def\SVars{{{\rm SVars}}}
\def\Val{{{\rm Val}}}
\def\MSO{{{\rm MSO}}}
\def\Sep{{{\rm sep}}}
\def\THeaps{{\rm THeaps}}
\def\Root{{\rm Root}}
\def\TG{{\rm TGraph}}

\def\Tilde{\widetilde}
\def\Bar{\overline}
\def\Lequiv{\Longleftrightarrow}
\def\Lto{\Longrightarrow}
\def\Lfrom{\Longleftarrow}
\def\Norm{{\rm Norm}}
\def\Noshare{{\rm Noshare}}
\def\Roots{{\rm Roots}}
\def\Forest{{\rm Forest}}
\def\Var{{\rm Var}}
\def\Range{{\rm Range}}
\def\Cell{{\rm Cell}}
\def\Tree{{\rm Tree}}
\def\Switch{{\rm Switch}}
\def\All{{\rm All}}
\def\To{\leadsto}
\def\tree{{\rm tree}}
\def\Paths{{\rm Paths}}
\def\Finite{{\rm Finite}}
\def\Const{{\rm Const}}
\def\Leaf{{\rm Leaf}}
\def\LeastElem{{\rm LeastElem}}
\def\LeastIndex{{\rm LeastIndex}}
\def\WSnS{{{\rm WSnS}}}
\def\Expand{{{\rm Expand}}}


\long\def\J#1{} 
\def\T#1{\hbox{\color{green}{$\clubsuit #1$}}}
\def\W#1{\hbox{\color{Orange}{$\spadesuit #1$}}}

\def\Node{{{\rm Node}}}
\def\LL{{{\rm LL}}}
\def\DSN{{{\rm DSN}}}
\def\DCL{{{\rm DCL}}}
\def\LS{{{\rm LS}}}
\def\Ls{{{\rm ls}}}

\def\FPV{{{\rm FPV}}}
\def\Lfp{{{\rm lfp}}}
\def\IsHeap{{{\rm IsHeap}}}

\def\Equiv{\quad \equiv\quad }
\def\Null{{{\rm null}}}
\def\Emp{{{\rm emp}}}
\def\If{{{\rm if\ }}}
\def\Then{{{\rm \ then\ }}}
\def\Else{{{\rm \ else\ }}}
\def\While{{{\rm while\ }}}
\def\Do{{{\rm \ do\ }}}
\def\Cons{{{\rm cons}}}
\def\Dispose{{{\rm dispose}}}
\def\Vars{{{\rm Vars}}}
\def\Locs{{{\rm Locs}}}
\def\States{{{\rm States}}}
\def\Heaps{{{\rm Heaps}}}
\def\FV{{\rm FV}}
\def\True{{{\rm true}}}
\def\False{{{\rm false}}}
\def\Dom{{{\rm Dom}}}
\def\Abort{{{\rm abort}}}
\def\New{{{\rm New}}}
\def\W{{{\rm W}}}
\def\Pair{{{\rm Pair}}}
\def\Lh{{{\rm Lh}}}
\def\lh{{{\rm lh}}}
\def\Elem{{{\rm Elem}}}
\def\EEval{{{\rm EEval}}}
\def\PEval{{{\rm PEval}}}
\def\HEval{{{\rm HEval}}}
\def\EVal{{{\rm Eval}}}
\def\Domain{{{\rm Domain}}}
\def\Exec{{{\rm Exec}}}
\def\Store{{{\rm Store}}}
\def\Storecode{{{\rm Storecode}}}
\def\Heapcode{{{\rm Heapcode}}}
\def\Lesslh{{{\rm Lesslh}}}
\def\Addseq{{{\rm Addseq}}}
\def\Separate{{{\rm Separate}}}
\def\Result{{{\rm Result}}}
\def\Lookup{{{\rm Lookup}}}
\def\ChangeStore{{{\rm ChangeStore}}}
\def\ChangeHeap{{{\rm ChangeHeap}}}
\def\Wand{{\mathbin{\hbox{\hbox{---}$*$}}}}
\def\Wando#1{{\mathop{\hbox{\hbox{---}$*$}}}^{#1}}
\def\Eval#1{[\kern -1pt[{#1}]\kern -1pt]}
\def\Vec{\overrightarrow}

\def\Tilde{\widetilde}
\def\Break{\hfil\break\hbox{}}

\title{Completeness of Cyclic Proofs \\ for Symbolic Heaps}

\author{
Makoto Tatsuta \\
\parbox{25ex}{National Institute of Informatics / Sokendai,\\ Japan} \\
{tatsuta@nii.ac.jp}
\and
Koji Nakazawa \\
Nagoya University, Japan \\
{knak@i.nagoya-u.ac.jp}
\and
Daisuke Kimura \\
Toho University, Japan \\
{kmr@is.sci.toho-u.ac.jp}
}

\date{}

\maketitle

\begin{abstract}

Separation logic is successful for software verification in both
theory and practice.  Decision procedure for symbolic heaps is one of
the key issues.  This paper proposes a cyclic proof system for
symbolic heaps with general form of inductive definitions, and shows
its soundness and completeness. The decision procedure for entailments
of symbolic heaps with inductive definitions is also given.
Decidability for entailments of symbolic heaps with inductive
definitions is an important question.  Completeness of cyclic proof
systems is also an important question.  The results of this paper
answer both questions.  The decision procedure is feasible since it is
nondeterministic double-exponential time complexity.

\end{abstract}

\section{Introduction}

Separation logic is successful for software verification \cite{Reynolds02,OHearn04,OHearn05}.
Several systems based on this idea have been actively investigated
and implemented.
One of the keys in these systems is the entailment checker that
decides the validity for a given entailment of symbolic heaps.

The paper \cite{Iosif13} proposed the system $\SLRDbtw$,
which is the first decidable system for
entailment of symbolic heaps with general form of inductive definitions.
We call the conditions 
imposed in \cite{Iosif13} for restricting the class of inductive definitions
by a {\em bounded treewidth condition}.
The inductive definitions without any restriction cause
undecidability \cite{Kanovich14}.
The bounded treewidth condition
is one of the most flexible conditions for a decidable system.

A cyclic proof system \cite{Brotherston11} can give us efficient implementation
of theorem provers.
On the other hand, the completeness and the decidability of provability
are not known for any cyclic proof system.
Hence it is a challenging problem 
to find some complete cyclic proof system and some decidable system for
symbolic heaps with general form of inductive definitions.

Our contribution is to solve these problems, namely,
to propose
a cyclic proof system $\CSLIDomega$
for symbolic heaps with inductive definitions,
to prove
its soundness theorem and
its completeness theorem, 
and
to give its decision procedure.

Our first ideas are as follows:
(1) We define inductive definition clauses so that the unfolding
determines the root cells of children.
(2) We use unfold-match and split,
namely, first unfold inductive predicates on both sides
(antecedent and succedent),
next remove the same $x \mapsto (\Underscore)$ on both sides,
then split the entailment by separating conjunction.
(3) We do proof-search by going up from the conclusion to the assumption 
in an inference rule.
We will show the termination by
defining normal forms
and
showing the set of those possibly used during computation is finite.
This shows the termination
since (a) every path of potentially infinite length 
contains infinite number of normal forms,
(b) the set of normal forms is finite,
(c) hence there is some repetition of normal forms in the path,
(d) hence in the path some subgoal of the repetition is eventually discharged
by cyclic proof mechanism.
(4) We also show (selective) local completeness of each application of rules
used in each step.
By this and termination, we will show the completeness.

Our main ideas are as follows.
Each of them is new.
All of these techniques together 
make an algorithm based on those first ideas a real algorithm.
(1) 
We will introduce 
atomic formulas $t \downarrow$ and $t \uparrow$ which mean
$t$ is in the heap and $t$ is not in the heap respectively,
and put $\downarrow$ or $\uparrow$ for each variable in
both antecedent and succedent.
(2) 
When we unfold the succedent, to keep validity, we need disjunction
in the succedent.
So we will introduce disjunction in the succedent.
(3) 
We will propose a new $(*)$-split rule for disjunction.
As far as we know, no $(*)$-split rule has not been proposed for disjunction.
(4) 
We will introduce a {\em factor} rule.
Roughly speaking, if the candidate of a common root is $x$ but it does not
appear in some disjunct $P(y)$,
we transform this disjunct into $(Q(x) \Wand P(y)) * Q(x)$
so that the disjunct has the root $x$.
(5) 
For splitting existential scopes,
we transform $\exists w((P(x,w) \land w\downarrow) * Q(y,w))$
into
$\exists w((P(x,w) \land w\downarrow) * (Q(y,w) \land w\Uparrow))$
and
it into
$\exists w((P(x,w) \land w\downarrow)) * \exists w((Q(y,w) \land w\Uparrow))$.
We will show these transformations keep equivalence.
(6) 
We eliminate a disjunct that is a renaming of another disjunct,
and moreover we will show that this elimination keeps validity.

For unfold-match and removing $\mapsto$,
we need some conditions (strong connectivity, decisiveness) to the class of inductive definitions
besides the bounded treewidth condition.
The establishment condition in the bounded treewidth condition
is checked by considering the set of inductive definitions,
and it is not locally checked by the shape of each definition clause.
Our condition is a local version of the bounded treewidth condition.
These additional conditions are not so restrictive
and our class of inductive definitions is still quite large, 
since our class contains
doubly-linked lists, skip lists, and nested lists. 

The decision procedure is feasible since it is double-exponential
time complexity.

Several entailment checkers for
symbolic heaps with inductive definitions
have been discussed.
Most of them do not have general form of inductive definitions
and have only hard-coded inductive predicates
\cite{Preze13,Piskac13,Piskac14,OHearn04,OHearn05,Cook11,Enea13,Enea14}.
The entailment checkers for general form of inductive definitions
are studied in \cite{Chin12,Chu15,Iosif13,Iosif14,Tatsuta15,Brotherston11b,Brotherston12}.
The engines of the system $\SLRDbtw$ \cite{Iosif13,Iosif14}
and the system in \cite{Tatsuta15}
are both model theoretic, and they are decidable systems.
The systems in \cite{Chu15,Brotherston11b,Brotherston12}
use cyclic proofs, but
neither of them is a complete system.
\cite{Chin12} is based on ordinary sequent calculus and is not complete.

The cyclic proofs have been intensively investigated
for the first-order predicate logic \cite{Brotherston11,Brotherston12,Berardi17,Simpson17,Berardi17a},
a bunched implication system \cite{Brotherston07},
and
a symbolic heap system \cite{Brotherston11b,Brotherston12}.

Section 2 defines separation logic with inductive definitions.
In Section 3,
we extend our language,
in particular,
we give definition clauses for the strong wand.
Section 4 defines the system $\CSLIDomega$.
Section 5 shows soundness of the factor rule.
Section 6 proves soundness of the existential amalgamation rules.
Section 7 proves soundness of the $(*)$-split rule.
Section 8 proves the soundness of the system.
Section 9 gives a satisfiability checker.
In Section 10, we define
normal forms and groups,
give the decision algorithm of validity,
and shows its partial correctness, loop invariants and termination.
Section 11 proves a property for constant store validity.
Section 12 explains cones.
Section 13 proves local completeness of the factor rule.
Section 14 shows local completeness of the unrelated introduction.
Section 15 shows selective local completeness of the rule $(*)$-split.
Section 16 shows properties for elimination of fresh variables in
the succedent.
Section 17 proves completeness of $\CSLIDomega$.
We conclude in Section 18.

\section{Symbolic Heaps with Inductive Definitions}

This section defines symbolic heaps, inductive definitions, and their semantics.

We will use vector notation $\Vec x$ to 
denote a sequence $x_1,\ldots,x_k$ for simplicity.
$|\Vec x|$ denotes the length of the sequence.
Sometimes we will also use a notation of a sequence to denote
a set for simplicity.
We write $\equiv$ for the syntactical equivalence.

\subsection{Language}

Our language is a first order logic with a new connective $*$ and
inductive predicates, and defined as follows.

First-order variables $\Vars ::= x,y,z,w,v, \ldots$.
\qquad
Terms $t,u,p,q,r ::= x \ |\  \Nil$.

Inductive Predicate Symbols $P,Q, R$.

We define formulas $F,G$ of separation logic as
those of the first-order language generated
by the constant $\Nil$, the propositional constant $\Emp$,
predicate symbols $=$, $\mapsto$, $P$,$Q$,$\ldots$,
and an additional logical connective $*$.
We write $t \ne u$ for $\neg t=u$.
We assume some number $\Ncell$ for the number of elements in a cell.

     Pure formulas $\Pi ::= t=t  \ |\  t \ne t  \ |\  \Pi \land \Pi$.

Spatial formulas $\Sigma ::= \Emp  \ |\  t \mapsto (t_1,\ldots,t_{\Ncell})  \ |\  P(\Vec t)  \ |\  \Sigma * \Sigma$.

We suppose $*$ binds more tightly than $\land$.
We will sometimes write $P(t)$ for $P(t,\Vec t)$.
We write $*_{i \in [1,n]} P_i(x_i)$ for $P_1(x_1) * \ldots * P_n(x_n)$.
Similarly we write $*_{i \in I}P_i(x_i)$.
We write $\Pi \subseteq \Pi'$ when
all the conjuncts of $\Pi$ are contained in those of $\Pi'$.

qf-Symbolic Heaps $A,B ::= \Pi \land \Sigma \ |\ \Sigma$.
\qquad
Symbolic Heaps $\phi ::= \exists\Vec xA$.

Entailments $A \prove B_1,\ldots,B_n$.

Inductive Definitions $P(x,\Vec y) \eqDef \Lor_i \phi_{i}(x,\Vec y)$
where $\phi_i$ is a definition clause.

Definition Clauses 
$
\phi(x,\Vec y) \equiv
\exists \Vec z (\Pi \land x \mapsto (\Vec u) *
*_{i \in I} P_{i}(z_i,\Vec t_i)),
$
where
\begin{itemize}

\item $\{z_i \ |\ i \in I \} = \Vec z$ (strong connectivity),

\item $\Vec z \subseteq \Vec u$ (decisiveness).

\end{itemize}

We call the first argument of a spatial atomic formula except $\Emp$ a {\em root}.

The strong connectivity implies the bounded treewidth condition.
The decisive condition is that
in every definition clause,
all existential variables must occur in $\Vec u$
where the definition clause has $x \mapsto (\Vec u)$.
It is similar to the constructively valued condition in \cite{Brotherston16}.
This condition guarantees that
the cell at address $x$ decides the content of every existential variable.

We give some examples of the inductive definitions in the following.

The list segment is definable:
$
\Ls(x,y) \eqDef x \mapsto y \lor \exists z(x \mapsto (z) * \Ls(z,y)).
$

The doubly-linked list is definable:

$
\Dll(h,p,n,t) \eqDef h=t \land h \mapsto (p,n) \lor \exists z(h \mapsto (p,z)*\Dll(z,h,n,t)).
$

The nested list is definable:

$
\Listnest(x) \eqDef \exists z.x \mapsto (z,\Nil)*\Ls(z,\Nil) \lor
	\exists z_1 z_2(x \mapsto (z_1,z_2)*\Ls(z_1,\Nil)*\Listnest(z_2)).
$

The following nested list segment is also definable.

$
\Lsnest(x,y) \eqDef \exists z(x \mapsto (z,\Nil)*\Ls(z,y)) \lor
	\exists z_1 z_2(x \mapsto (z_1,z_2)*\Ls(z_1,y)*\Lsnest(z_2,y)).
$

The skip list is definable:

$
\Skl1(x,y) \eqDef x \mapsto (\Nil,y) \lor \exists z(x \mapsto (\Nil,z)*\Skl1(z,y)),$

$
\Skl2(x,y) \eqDef \exists z(x \mapsto (y,z) * \Skl1(z,y)) \lor
  \exists z_1 z_2( x \mapsto (z_1,z_2)*\Skl1(z_2,z_1)*\Skl2(z_1,y)).
$

The examples in \cite{Brotherston11b} are definable in our system as follows:
List, ListE, ListO are definable,
RList is not definable.
DLL, 
PeList, SLL, BSLL, BinTree, BinTreeSeg, BinListFirst, BinListSecond, BinPath
are not definable but will be definable in a straightforward extension
of our system by handling $\Emp$ in the base cases.

We prepare some notions.
We define $P^{(m)}$ by
\[
P^{(0)}(\Vec x) \equiv (\Nil \ne \Nil), \\
P^{(m+1)}(\Vec x) \equiv \Lor_i \phi_i[P:=P^{(m)}],
\]
where $P(\Vec x) \eqDef \Lor_i\phi_i$.
$P^{(m)}$ is $m$-time unfold of $P$.
We define $F^{(m)}$ as obtained from a formula $F$ by replacing every inductive predicate $P$ by $P^{(m)}$.

We define $(\ne(T_1,T_2))$ as 
$\Land_{t_1 \in T_1,t_2 \in T_2, t_1\not\equiv t_2}t_1\ne t_2$.
We write $x \ne T$ for $(\ne(\{x\},T))$.

We define $(\ne(T))$ as $(\ne(T \cup \{\Nil\},T \cup \{\Nil\}))$.

\subsection{Semantics}

This subsection gives semantics of the language.

We define the following structure:
$\Val = N$,
$\Locs = \{ x \in N | x > 0 \}$,
$\Heaps = \Locs \to_{fin} \Val^\Ncell$,
$\Stores = \Vars \to \Val$.
Each $s \in \Stores$ is called a {\it store}. 
Each $h \in \Heaps$ is called a {\it heap}, and 
$\Dom(h)$ is the domain of $h$,
and 
$\Range(h)$ is the range of $h$.
We write $h=h_1+h_2$ when $\Dom(h_1)$ and $\Dom(h_2)$ are
disjoint and the graph of $h$ is the union of those of $h_1$ and $h_2$.
A pair $(s,h)$ is called a {\it heap model}, which means a memory state.
The value $s(x)$ means the value of the variable $x$ in the model $(s,h)$. 
Each value $a \in \Dom(h)$ means an address, and 
the value of $h(a)$ is the content of the memory cell at address $a$ 
in the heap $h$. 
We suppose each memory cell has $\Ncell$ elements as its content.

The interpretation $s(t)$ for any term $t$ is defined 
as $0$ for $\Nil$ and $s(x)$ for the variable $x$. 

For a formula $F$
we define the interpretation $s,h \models F$ 
as follows. 

$s,h \models t_1 = t_2$ if $s(t_1) = s(t_2)$, 

$s,h \models F_1 \land F_2$ if $s,h \models F_1$ and $s,h \models F_2$, 

$s,h \models F_1 \lor F_2$ if $s,h \models F_1$ or $s,h \models F_2$, 

$s,h \models \neg F$ if $s,h \not\models F$.

$s,h \models \Emp$ if $\Dom(h)=\emptyset$, 

$s,h \models t \mapsto (t_1,\ldots,t_\Ncell)$ if $\Dom(h) = \{ s(t) \}$ and 
$h(s(t)) = (s(t_1),\ldots,s(t_\Ncell))$, 

$s,h \models F_1  * F_2$ 
if $s,h_1 \models F_1$ and
$s,h_2 \models F_2$ for some $h_1 + h_2=h$, 

$s,h \models P(\Vec t)$ if $s,h \models P^{(m)}(\Vec t)$ for some $m$,

$s,h \models \exists zF$ if $s[z:=b],h \models F$ for some $b \in \Val$. 

We write
$A \models B_1,\ldots, B_n$ for
$\forall sh(s,h \models A \imp ((s,h \models B_1) \lor \ldots \lor (s,h \models B_n)))$.
The entailment $A \vdash B_1,\ldots,B_n$ 
is said to be valid if $A \models B_1,\ldots,B_n$ holds. 
Our goal in this paper is to decide the validity of a given entailment.

For saving space, we identify some syntactical objects that have the same
meaning, namely,
we use implicit transformation of formulas by using the following properties:
$\land$ is commutative, associative, and idempotent;
$*$ is commutative, associative, and has the unit $\Emp$;
$=$ is symmetric;
$\Pi \land (F*G) \lequiv (\Pi \land F)*G$;
$\exists xG \lequiv G$,
$\exists x(F \land G) \lequiv \exists xF \land G$,
and
$\exists x(F * G) \lequiv \exists xF * G$,
when $F,G$ are formulas and $x \notin \FV(G)$;
$\exists xyF \lequiv \exists yxF$.
We also identify the succedent of an entailment with a set of disjuncts
instead of a sequence of disjuncts.

\section{Language Extension}

\subsection{Extended Language}

In this section, we extend our language from symbolic heaps
by $\downarrow$ and $L$ and $\Swand$,
which is necessary to show the completeness.

We extend inductive predicate symbols with
$Q_1 \Swand \ldots \Swand Q_m \Swand P$
where $Q_1,\ldots,Q_m,P$ are original inductive predicate symbols.
We call $m$ the {\em depth of wands}.
We write
$Q_1(\Vec t_1) \Swand \ldots \Swand Q_m(\Vec t_m) \Swand P(\Vec t)$
for
$(Q_1 \Swand \ldots \Swand Q_m \Swand P)(\Vec t,\Vec t_1,\ldots,\Vec t_m)$.
For a sequence $\vec{R}
= R_1,\cdots,R_n$ of predicates, we write $\vec{R}\Wand P$ for $R_1
\Wand \cdots \Wand R_n \Wand P$.

We extend our first-order language with 
the extended inductive predicate symbols and 
unary predicate symbols $\downarrow$ and $L$.
$t \downarrow$ means that $t$ is in $\Dom(h)$ and
$L(t)$ means that $t$ is in $\Range(h)-\Dom(h)$
(the leaves of $h$).
We write $t \uparrow$ for $\neg t \downarrow$.
We write $t\Uparrow$ for $t\uparrow \land \neg Lt$.

We write $\Sigma, A,B,\phi$ for
the same syntactical objects with the extended inductive predicate symbols.

We use $X,Y$ for a finite set of variables and
write $X \uparrow$ for
$\Land \{ t \uparrow \ |\ t \in X \}$.
$X\downarrow$ and $X\Uparrow$ are similarly defined.
We write $\exists\Vec x\downarrow$ for $\exists\Vec x(\Vec x\downarrow \land \ldots)$. Similarly we write $\exists\Vec x\Uparrow$.

We define:

$\calP ::= \ \mapsto  \ |\ P$ where $P$ varies in inductive predicate symbols,

$\Delta ::= \calP(\Vec t) \land X \downarrow$,
and
$\Gamma ::= \Delta \ |\ \Gamma * \Gamma$,
and
$\psi ::= Y \uparrow \land \Pi \land \Gamma$,
and
$\Phi ::= \exists \Vec x\exists\Vec y\Uparrow(\Pi \land \Gamma)$,

$F,G$ a separation logic formula with $\downarrow,L$.

We define  entailments as 
$\psi \prove \Phi_1,\ldots,\Phi_n$.

We write $J$ for an entailment.
In $\psi, \Phi$,
we call $\Gamma$ a {\em spatial part} and $\Pi$ a {\em pure part}.

We define 
$
\Roots(X\Uparrow \land Y \uparrow \land \Pi \land *_{i \in I} (\calP_i(x_i,\Vec t_i) \land X_i\downarrow)) = \{ x_i | i \in I \}
$.
Then we define
$\Roots(\exists x \Phi) = \Roots(\Phi)$ if $x \notin \Roots(\Phi)$,
and undefined otherwise.
We define
$
\Cells(X\Uparrow \land Y\uparrow \land \Pi \land 
*_{i \in I} (\calP_i(x_i) \land X_i \downarrow)) = \bigcup_{i \in I } X_i.
$
Then we define $\Cells(\exists x\Phi) = \Cells(\Phi)-\{x\}$.

We write $(\Roots+\Cells)(F)$ for $\Roots(F)+\Cells(F)$
and call them {\em address variables} of $F$.

We define a substitution as a map from
the set of variables to the set of terms.
For a substitution $\theta$, 
we define 
$\Dom(\theta) = \{ x | \theta(x) \not\equiv x \}$
and
$\Range(\theta) = \{ \theta(x) | x \in \Dom(\theta) \}$.
We define a {\em variable renaming} as
a substitution that is a bijection among variables with a finite domain.

\begin{Def}\rm
$s,h \models t \downarrow$ iff $s(t) \in \Dom(h)$.

$s,h \models L(t) $ iff $s(t) \in \Range(h)-\Dom(h)$.
\end{Def}

For saving space, we identify some syntactical objects that have the same
meaning, namely,
we use implicit transformation of formulas by using the following properties:
$(F*G) \land X\uparrow \lequiv (F \land X\uparrow)*(G \land X\uparrow)$;
$(F*G) \land X\Uparrow \lequiv (F \land X\Uparrow)*(G \land X\Uparrow)$.

\subsection{Strong Wand}

This section gives
the definition clauses for inductive predicates that contain the strong wand.

\begin{Def}\rm
The definition clauses of $Q(y,\Vec w) \Swand P(x,\Vec y)$
are as follows:

Case 1.
$
\exists (\Vec z-z_i)((\Vec w = \Vec t_i
\land
\Pi \land x \mapsto (\Vec u) * *_{l \ne i} P_l(z_l,\Vec t_l))[z_i:=y])
$
where
$Q = P_i$ and
$
\exists \Vec z(\Pi \land x \mapsto (\Vec u) * *_l P_l(z_l,\Vec t_l))
$
is a definition clause of $P(x,\Vec y)$.

Case 2.
$
\exists \Vec z(\Pi \land x \mapsto (\Vec u) * *_{l \ne i, l \in L} P_l(z_l,\Vec t_l) *
(Q(y,\Vec w) \Swand P_i(z_i,\Vec t_i)))
$
where $i \in L$ and
$
\exists \Vec z(\Pi \land x \mapsto (\Vec u) * *_{l \in L} P_l(z_l,\Vec t_l))
$
is a definition clause of $P(x,\Vec y)$.
\end{Def}
$Q(y) \Swand P(x)$ is inductively defined by 
the definition clauses obtained by removing $Q(y)$ from
the definition clauses of $P(x)$. 
$Q(y) \Swand P(x)$ plays a similar role to
the ordinary magic wand $Q(y) \Wand P(x)$,
but it is stronger than the ordinary magic wand and it is defined syntactically.
Roughly speaking, it is defined to be false if
it cannot be defined syntactically.

\begin{Eg}
$
\Ls(y,v) \Swand \Ls(x,w) \eqDef w = v \land x \mapsto (y) \lor
\exists z(x \mapsto (z) * (\Ls(y,v) \Swand \Ls(z,w))).
$
\end{Eg}

Note that $Q_1(\Vec t_1) \Swand Q_2(\Vec t_2) \Swand P(\Vec t)$ and
$Q_2(\Vec t_2) \Swand Q_1(\Vec t_1) \Swand P(\Vec t)$ are
equivalent, which will be shown by Lemma \ref{lemma:multiwand}.

\section{System $\CSLIDomega$}

This section defines our logical system $\CSLIDomega$.

\subsection{Inference Rules}

This subsection gives the set of inference rules.

We define $\Dep(P)$ as the set of inductive predicates symbols
that appear in the unfolding of $P$.
We write $F[F']$ to explicitly display the subformula $F'$ in $F$.
We write $T$ for a finite set of terms.
We say $\Phi$ is {\em equality-full} when
$\Pi$ contains $(\ne(\Vec y,V \cup \Vec y \cup \{\Nil\}))$
where
$\Phi$ is $\exists \Vec x\exists\Vec y\Uparrow(\Pi \land \Gamma)$
and 
$V = \FV(\Phi)$.
We write $\Pi' \subseteq \exists \Vec x\exists\Vec y\Uparrow(\Pi \land \Gamma)$
when $\Pi' \subseteq \Pi$.
We call a context $F[\ ]$ {\em positive} when
the number of $\neg(\ )$ and $(\ )\imp F'$ that contain $[\ ]$ is even.
We call a context $F[\ ]$ {\em existential} when
it is of the form $\exists\Vec x([\ ] * F')$.

We assume a number $\Dw$ for the maximum depth of wands.

The inference rules are given in the figure \ref{fig:rules}.
\begin{figure}[]
{\prooflineskip \small
\[
\infer[(\Subst)]{F \theta \prove \Vec G \theta}{
F \prove \Vec G
}
\qquad
 \infer[(\Emp)]{\Emp \prove \Emp}{}
\qquad
\infer[(\Unsat)]{F \prove \Vec G}{}
\ \ (F {\rm\ unsatisfiable})
\\
\infer[(\land L)]{F' \land F \prove \Vec G}{
F \prove \Vec G
}
\qquad
\infer[(\land\Elim)]{F \prove \Vec G, \exists\Vec xG}{
F \prove \Vec G, \exists\Vec x(G \land G')
}
\qquad
\infer[(\lor\Elim)]{F \prove \Vec G}{
F \prove \Vec G, \exists\Vec x(G \land \neg G)
}
\\
\infer[(\land R)]{F \land F' \prove \Vec G, G \land F'}{
F \land F' \prove \Vec G, G
}
\qquad
\infer[(\uparrow\Elim)]{F_1 * F_2 \prove \Vec G}{
(x \uparrow * F_1) * F_2 \prove \Vec G
}
\ (x \in (\Roots+\Cells)(F_2))
\\
\infer[(\lor R)]{F \prove \Vec G, G}{ 
 F \prove \Vec G}{
 }
\qquad
\infer[(\exists R)]{F \prove \Vec G, \exists wG}{
F \prove \Vec G, G[w:=t]
}
\qquad
 \infer[(=L)]{x=t \land F \prove \Vec G}{
F[x:=t] \prove \Vec G[x:=t]
}
\qquad
\infer[(=R)]{F \prove \Vec G, G \land t=t}{
F \prove \Vec G,G
}
\\
\infer[(\downarrow\Out\ L)]{F*((F_1 * F_2) \land x\downarrow) \prove \Vec G}{
F*(F_1 \land x\downarrow) * F_2 \prove \Vec G
&
F*F_1 * (F_2 \land x\downarrow) \prove \Vec G
}
\qquad
\infer[(\Uparrow R)]{F \prove \Vec G,\exists y\Uparrow(y \ne T \land G)}{
F \prove \Vec G,G
}
\\
\infer[(\downarrow \Out\ R)]{F \prove \Vec G, \exists\Vec x(G * ((G_1 * G_2) \land t \downarrow)) }{
F \prove \Vec G, \exists\Vec x(G * (G_1 \land t \downarrow) * G_2),
\exists\Vec x(G * G_1 * (G_2 \land t \downarrow))
}
\qquad
\infer[(\downarrow R)]{F * \calP(t,\Vec t) \prove \Vec G, G \land t\downarrow}{
F * \calP(t,\Vec t) \prove \Vec G, G
}
\\
\infer[(\Case\ L)]{F \prove \Vec G}{
F \land F' \prove \Vec G
&
F \land \neg F' \prove \Vec G
}
\\
\infer[(\Factor)]{F \prove \Vec G,
G[\Vec {Q_1(\Vec t_1)},\Vec {Q_2(\Vec t_2)} \Swand P(\Vec t)]}{
\hbox to 80ex{\vbox{\noindent $
F \prove \Vec G,
\{G[
\exists\Vec w(
(\Vec {Q_1(\Vec t_1)},Q(y,\Vec w) \Swand P(\Vec t)) *
(\Vec {Q_2(\Vec t_2)} \Swand Q(y,\Vec w))
)
]
\ |\
$
\hfil\break\qquad $
Q \in \Dep(P), \Vec Q_2 \subseteq \Dep(Q),
\Vec w {\rm\ fresh}\}
$}}
}
\ (G[\ ] {\rm\ existential})
\\
\infer[(\Pred\ L)]{P(x,\Vec t) * F \prove \Vec G}{
A(x,\Vec t,\Vec z) * F \prove \Vec G
&
\hbox{(for every definition clause $\exists\Vec z A(x,\Vec t,\Vec z)$ of $P(x,\Vec t)$)}
}
\\
 \infer[(\Pred\ R)]{F \prove \Vec G, G[P(x,\Vec t)]}{
F \prove \Vec G,
\{
 G[\phi] \ |\
 \hbox{$\phi$ is a definition clause of $P(x,\Vec t)$}
\}
}
\ (G[\ ] {\rm\ existential})
\\
\infer[(*\mapsto)]{F * x \mapsto (\Vec u) \prove
	\{ G_i * x \mapsto (\Vec u) \ |\ i \in I \}}{
x \uparrow \land F \prove
	\{ G_i \ |\ i \in I \}
}
\\
\infer[(\exists\ \Amalgamation1)]{F \prove \Vec G,
G[\exists x\Uparrow(\Phi_1 * \Phi_2)]
}{
F \prove \Vec G,
G[\exists x\Uparrow\Phi_1 *
\exists x\Uparrow\Phi_2]
}
(\hbox{$\exists x\Uparrow\Phi_1$, $\exists x\Uparrow\Phi_2$ equality-full, 
$G[\ ]$ positive})
\\
\infer[(\exists\ \Amalgamation2)]{F \prove \Vec G,
G[\exists x(\Phi_1 * (x\Uparrow \land \Phi_2))]
}{
F \prove \Vec G,
G[\exists x\Phi_1 *
\exists x\Uparrow\Phi_2]
}
\left(\vbox{\hbox{$x \in \Cells(\Phi_1)$,
$(x \ne \FV(\exists x\Uparrow\Phi_2)) \subseteq \Phi_1$,}
\hbox{$\exists x\Uparrow\Phi_2$ equality-full, $G[\ ]$ positive}}
\right)
\\
\infer[(*)]{F_1 * F_2 \prove \{ G_1^i * G_2^i \ |\  i \in I \} }{
F_1 \prove \{G_1^i \ |\ i \in I' \}
{\rm\ or\ }
F_2 \prove \{ G_2^i \ |\ i \in I-I' \}
&
(\forall I' \subseteq I)
}
\]
}
\caption{Inference Rules}\label{fig:rules}
\end{figure}

The rule (\Factor) derives $P(t)$ from $(Q(y) \Swand P(t))*Q(y)$ where
we list up all the cases for $Q(y)$ in the disjunction.
The rules $(\exists\ \Amalgamation1,2)$ amalgamate 
$\exists x$'s under some condition,
which guarantees that existentials have the same values.
The rule $(*)$ is a new split rule since it handles disjunction in the succedent.
The other rules are standard.

\subsection{Proofs in $\CSLIDomega$}

This subsection defines a proof in $\CSLIDomega$.

\begin{Def}\rm
We define a {\em bud} and a {\em companion} in the same way as
\cite{Brotherston11}.
For $\CSLIDomega$, we define a {\em cyclic proof} to be
a proof figure by the inference rules
without any open assumptions
where
each bud has a companion below it
and there is some rule $(*\mapsto)$ between them.
\end{Def}

Instead of the global trace condition 
in ordinary cyclic proof systems \cite{Brotherston11},
$\CSLIDomega$
requires some $(*\mapsto)$ rule between 
a bud and its companion.

\subsection{Preparation for Soundness Proof}

The following is a key notion for simple soundness proof.
\begin{Def}\rm
We write $F \models_m \Vec G$ when
for all $s,h$, $|\Dom(h)| \le m$ and $s,h \models F$
imply $s,h \models \Vec G$.
We say $F \prove \Vec G$ is {\em $m$-valid} and write
$\models_m F \prove \Vec G$
when $F \models_m \Vec G$.

We call a rule $m$-sound when the following holds:
if all the assumptions are $m$-valid  then 
the conclusion is $m$-valid.
\end{Def}

\begin{Lemma}\label{lemma:existsUparrow}
If $x \notin T$, then
\[
F \imp \exists x\Uparrow((x \ne T) \land F).
\]
\end{Lemma}

{\em Proof.}
Assume
\[
s,h \models F.
\]
There is $a$ such that $a \ne s(T)$ and $ \notin \Dom(h) \cup \Range(h)$
since the set of addresses is infinite.
Then
\[
s[x:=a],h \models x\Uparrow \land (x \ne T).
\]
Hence
\[
s,h \models \exists x\Uparrow((x \ne T) \land F).
\]
$\Box$

\begin{Def}\rm
For a heap $h$,
$\Leaves(h)$ is defined as $\Range(h)-\Dom(h)$.
\end{Def}

\begin{Lemma}\label{lemma:leaves}
If
\[
s,h \models \Gamma
\]
then 
$\Leaves(h) \subseteq \bigcup \{ s(\Vec t \cup \{\Nil\}) \ |\ P(x,\Vec t) \in \Gamma\}$.
\end{Lemma}

{\em Proof.}
By induction on $|h|$.

Let $\Gamma$ be $*_i \Delta_i$.
Assume $a \in \Leaves(h)$.
We have $h=\Sigma_i h_i$ such that $s,h_i \models \Delta_i$.
We have $i$ such that $a \in \Leaves(h_i)$.
Let $\Delta_i$ be $P(x,\Vec t) \land X\downarrow$.
Let
\[
s[\Vec z:=\Vec r],h \models \Pi \land x \mapsto (\Vec u) * *_l P_l(z_l,\Vec t_l).
\]
$u_i$ is included in $x,\Vec t,\Nil,\Vec z$.
Hence $s(\Vec u) - \Dom(h)$ is included in $s(\Vec t, \Nil)$.

Let $s'$ be $s[\Vec z:=\Vec r]$ and $h_i = h'+h''$ such that
\[
s',h' \models x \mapsto (\Vec u), \\
s',h'' \models *_l P_l(z_l,\Vec t_l).
\]
By IH, $\Leaves(h'') \subseteq \bigcup\{ s'(\Vec t_l \cup \{ \Nil \}) \ |\ l\}$.
Since $\Vec t_l$ are included in $x,\Vec t,\Nil,\Vec z$,
$s'(\Vec t_l)-\Dom(h)$ are included in $s(\Vec t,\Nil)$.

Hence $a$ is included in $s(\Vec t,\Nil)$.
Hence
$\Leaves(h) \subseteq \bigcup \{ s(\Vec t \cup \{\Nil\}) \ |\ P(x,\Vec t) \in \Gamma\}$.
$\Box$

\section{Soundness of Rule $(\Factor)$ and Properties for Strong Wands}

\def\vec{\Vec}

We prove some properties for strong wands. 

The order in $\vec{R}$ is not important in $\vec{R}\Swand P$,
since
$\vec{R}\Swand P$ and $\vec{R}'\Swand P$ are equivalent when $\vec{R}'$ is
a permutation of $\vec{R}$.
This is shown in the next lemma.
\begin{Lemma}\label{lemma:multiwand}
 The definition clauses of $\vec{R(v,\vec{u})} \Swand P(x,\vec{z})$
 are 
\[
  \exists(\vec{w}_{1i})_{i\in I}(\Pi\land \bigwedge
 \vec{\vec{u}_2=\vec{s}_2} \land x\mapsto (\vec{t}) *
 \ast_{i\in I}(\vec{R_{1i}(v_{1i},\vec{u}_{1i})} \Swand
 Q_{1i}(w_{1i},\vec{s}_{1i})))[\vec{w}_2:=\vec{v}_2]\quad (\#)
\]
for each definition clause $\exists\vec{w}(\Pi \land x\mapsto (\vec{t})
* \vec{Q(w,\vec{s})})$ of $P(x,\vec{z})$, and divisions
$\vec{R(v,\vec{u})}=(\vec{R_{1i}(v_{1i},\vec{u}_{1i})})_{i\in
I},\vec{R_2(v_2,\vec{u}_2)}$ (with some permutation) and
$\vec{Q(w,\vec{s})}=(Q_{1i}(w_{1i},\vec{s}_{1i}))_{i\in
 I},\vec{Q_2(w_2,\vec{s}_2)}$ such that $\vec{R}_2=\vec{Q}_2$.
 Hence, if $\vec{R}'$ is a permutation of $\vec{R}$, then $\vec{R}'\Swand
P$ is equivalent to $\vec{R}\Swand P$. Note that each
$\vec{R}_{1i}$ can be empty.
\end{Lemma}

{\em Proof.}
By induction on the length of $\vec{R}$.

Case $0$.
In this case, $\vec{Q} = (Q_{1i})_i$ and all of $(\vec{R}_{1i})_i$ and
$\vec{R}_2$ are empty. Hence,
$(\#)$ is the same as $\exists\vec{w}(\Pi \land x\mapsto (\vec{t})
* \vec{Q(w,\vec{s})})$.

Case $R',\vec{R}\Swand P(=R'\Swand(\vec{R}\Swand P))$. By the induction
hypothesis, a definition clause of $\vec{R}\Swand P$ is
\[
  \exists(\vec{w}_{1i})_{i\in I}(\Pi\land \bigwedge
 \vec{\vec{u}_2=\vec{s}_2} \land x\mapsto (\vec{t}) *
 \ast_{i\in I}(\vec{R_{1i}(v_{1i},\vec{u}_{1i})} \Swand
 Q_{1i}(w_{1i},\vec{s}_{1i})))[\vec{w}_2:=\vec{v}_2]
\]
for a definition clause $\exists\vec{w}(\Pi \land x\mapsto (\vec{t}) *
\vec{Q(w,\vec{s})})$ of $P$, and some divisions
$\vec{R(v,\vec{u})}=(\vec{R_{1i}(v_{1i},\vec{u}_{1i})})_{i\in
I},\vec{R_2(v_2,\vec{u}_2)}$ and
$\vec{Q(w,\vec{s})}=(Q_{1i}(w_{1i},\vec{s}_{1i}))_{i\in
I},\vec{Q_2(w_2,\vec{s}_2)}$ such that $\vec{R}_2=\vec{Q}_2$.

By the definition of the strong wand, we have two cases for the
definition clauses of $R'\Swand(\vec{R}\Swand P)$.

Subcase 1. $R'= \vec{R}_{1j}\Swand Q_{1j}$ for some $j$, that is,
$\vec{R}_{1j}$ is empty and $R' = Q_{1j}$, and the clause is
\[
  \exists(\vec{w}_{1i})_{i\ne j}(\Pi\land \bigwedge
 \vec{\vec{u}_2=\vec{s}_2}\land u_{1j}=s_{1j} \land x\mapsto (\vec{t}) *\\
 \qquad
 \ast_{i\ne j}(\vec{R_{1i}(v_{1i},\vec{u}_{1i})} \Swand
 Q_{1i}(w_{1i},\vec{s}_{1i})))[\vec{w}_2:=\vec{v}_2,w_{1j}:=v_{1j}].
\]
This is $(\#)$ for the division $R',\vec{R} = ((\vec{R}_{1i})_{i\ne
j},(R',\vec{R}_2))$.

Subcase 2. The clause is
\[
  \exists(\vec{w}_{1i})_i(\Pi\land \bigwedge
 \vec{\vec{u}_2=\vec{s}_2} \land x\mapsto (\vec{t}) *\\
 \qquad
 \ast_{i\ne j}(\vec{R_{1i}(v_{1i},\vec{u}_{1i})} \Swand
 Q_{1i}(w_{1i},\vec{s}_{1i}))
 \ast (R',\vec{R}_{1j}\Swand Q_{1j}))[\vec{w}_2:=\vec{v}_2,w_{1j}:=v_{1j}].
\]
This is $(\#)$ for the division $R',\vec{R} = ((\vec{R}'_{1i})_i,
\vec{R}_2)$, where $\vec{R}'_{1i} = \vec{R}_{1i}$ for $i\ne j$, and
$\vec{R}'_{1i} = R',\vec{R}_{1i}$ for $i=j$.
$\Box$

The following shows what is derived from the strong wand.

\begin{Lemma}[Strong Wand Elimination]\label{lemma:wand1}
$(Q(y,\vec{w})\Swand P(x,\vec{z})) *
(\vec{R(v,\vec{u})}\Swand Q(y,\vec{w}))
\models \vec{R(v,\vec{u})}\Swand P(x,\vec{z})$.
\end{Lemma}

{\em Proof.}
 We prove $s,h\models LHS$ implies $s,h\models RHS$ by induction on the
 size of $h$.
 
 Suppose $s,h\models LHS$, then we have some $h_1 + h_2 = h$ such that
 \[
 s,h_1 \models Q(y,\vec{w})\Swand P(x,\vec{z})
 \\
 s,h_2 \models \vec{R(v,\vec{u})}\Swand Q(y,\vec{w}).
 \]
By the definition of the strong wand, we have the following two cases.

Case 1. We have
 \[
 s,h_1\models \exists(\vec{r})(\vec{w}=\vec{s}_Q\land\Pi\land x\mapsto(\vec{u}) * *_l S_l(r_l,\vec{t}_l))[r_Q:=y]
 \]
for a definition clause 
$\exists(\vec{r}r_Q)(\Pi\land x\mapsto(\vec{t})* Q(r_Q,\vec{s}_Q) * *_l S_l(r_l,\vec{s}_l))$
of $P(x,\vec{z})$.
  Then, we have some $\vec{b}$ such that 
  \[
  s[\vec{r}:= \vec{b},r_Q:=s(y)],h_1 \models
\vec{w}=\vec{s}_Q\land\Pi\land x\mapsto(\vec{u}) * *_l S_l(r_l,\vec{t}_l).
  \]
  Let $s' = s[\vec{r}:= \vec{b},r_Q:=s(y)]$. We also have
  \[
   s',h_2 \models \vec{R(v,\vec{u})}\Swand Q(y,\vec{w}),
  \]
  since none of $\vec{r},r_Q$ occurs in $\vec{R(v,\vec{u})}\Swand
  Q(y,\vec{w})$. Hence we have
  \[
  s',h \models
   \Pi \land x\mapsto(\vec{t}) *
  (\vec{R(v,\vec{u})} \Swand Q(y,\vec{w})) * *_l S_l(r_l,\vec{t}_l).
  \]
  Since $s',h\models \vec{w}=\vec{s}_Q$ and $s'(r_Q)=s(y)$, we have
  \[
  s',h \models
   \Pi \land x\mapsto(\vec{t}) *
  (\vec{R(v,\vec{u})} \Swand Q(r_Q,\vec{s}_Q)) * *_l S_l(r_l,\vec{t}_l),
  \]
  and hence,
  \[
  s,h \models
   \exists(\vec{r}r_Q)(\Pi \land x\mapsto(\vec{t}) *
  (\vec{R(v,\vec{u})} \Swand Q(r_Q,\vec{s}_Q)) * *_l S_l(r_l,\vec{t}_l)),
  \]
  which is a definition clause of $\vec{R(v,\vec{u})}\Swand
  P(x,\vec{z})$ by Lemma \ref{lemma:multiwand}.

  Case 2. We have
  \[
  s,h_1\models \exists(\vec{r}r)(\Pi\land x\mapsto(\vec{u}) *
  (Q(y,\vec{w})\Swand S(r,\vec{s}))*_l S_l(r_l,\vec{t}_l))
  \]
  for a definition clause 
  $\exists(\vec{r}r)(\Pi\land x\mapsto(\vec{t})* S(r,\vec{s}) * *_l S_l(r_l,\vec{s}_l))$
  of $P(x,\vec{z})$.
  Then, we have some $\vec{b}$ and $b$ such that
  \[
   s[\vec{r}:=\vec{b},r:=b],h_1\models
   \Pi\land x\mapsto(\vec{u}) *
   (Q(y,\vec{w})\Swand S(r,\vec{s}))*_l S_l(r_l,\vec{t}_l).
  \]
  Let $s'=s[\vec{r}:=\vec{b},r:=b]$, and we have some
  $h_1=h_x+h_S+\Sigma_l h_l$ such that
  \[
  s',h_x \models x \mapsto (\vec{t})\\
  s',h_S \models Q(y,\vec{w})\Swand S(r,\vec{s})\\
  s',h_l \models 
  S_l(r_l,\vec{s}_l) \quad\mbox{(for each $l$)}.
  \]
  We also have
  \[
   s',h_2 \models \vec{R(v,\vec{u})}\Swand Q(y,\vec{w}),
  \]
  since none of $\vec{r},r$ occurs in $\vec{R(v,\vec{u})}\Swand Q(y,\vec{w})$.
  Therefore, we have
  \[
  s',h_S+h_2 \models 
  \vec{R(v,\vec{u})} \Swand S(r,\vec{s})
  \]
  by IH. Hence, we have
  \[
  s',h \models \Pi \land
  x\mapsto (\vec{t}) *
  (\vec{R(v,\vec{u})} \Swand S(r,\vec{s}))
  * *_l
  S_l(r_l,\vec{s}_l),
  \]
  and then
  \[
  s,h \models
  \exists(\vec{r}r)(\Pi \land
  x\mapsto (\vec{t}) *
  (\vec{R(v,\vec{u})} \Swand S(r,\vec{s}))
  * *_l
  S_l(r_l,\vec{s}_l)),
  \]
    which is a definition clause of $\vec{R(v,\vec{u})}\Swand
    P(x,\vec{z})$ by Lemma \ref{lemma:multiwand}.    
$\Box$

By the previous two lemmas, we have
\[
 (Q(y,\vec{w}),\vec{R(v,\vec{u})}\Swand P(x,\vec{z})) *
(\vec{R'(v',\vec{u}')}\Swand Q(y,\vec{w}))
\models \vec{R(v,\vec{u})},\vec{R'(v',\vec{u}')}\Swand P(x,\vec{z}).
\]

The following shows what derives the strong wand.
\begin{Lemma}[Strong Wand Introduction]\label{lemma:wand2}
  $x\ne y \land y \downarrow \land
 (\vec{R(v,\vec{u})}\Swand P(x,\vec{z}))
 \models
 \{\exists \vec{w}((Q(y,\vec{w}),\vec{R_1(v_1,\vec{u}_1)}\Swand
 P(x,\vec{z})) * (\vec{R_2(v_2,\vec{u}_2)}\Swand Q(y,\vec{w}))
 \mid
 \vec{R}=(\vec{R}_1,\vec{R}_2), Q\in\Dep(P), \vec{R}_2\subseteq \Dep(Q)
 \}$
\end{Lemma}

{\em Proof.}
 We prove $s,h\models LHS$ implies $s,h\models RHS$ by induction on the
 size of $h$.

  Suppose $s,h\models LHS$. Since $s,h\models
  \vec{R(v,\vec{u})}\Swand P(x,\vec{z})$, we have
  \[
  s,h\models \exists(r_{Ai})_i(\Pi\land
  \bigwedge\vec{\vec{u}_B=\vec{s}_B} \land x\mapsto(\vec{t}) *
  \ast_i (\vec{R_{Ai}(v_{Ai},\vec{u}_{Ai})} \Swand
  S_{Ai}(r_{Ai},\vec{s}_{Ai})))[\vec{r}_B:=\vec{v}_B]
  \]
  for some definition clause
  \[
  \exists(r_{Ai})_i\vec{r}_B(\Pi \land x\mapsto(\vec{t}) *
  \ast_i S_{Ai}(r_{Ai},\vec{s}_{Ai}) * \vec{S_B(r_B,\vec{s}_B)})
  \]
  of $P$ and some division
  $\vec{R(v,\vec{u})}=(\vec{R_{Ai}(v_{Ai},\vec{u}_{Ai})})_i,
  \vec{R_B(v_B,\vec{u}_B)}$ such that $\vec{R}_B=\vec{S}_B$.

  Then, we have some $(b_{Ai})_i$ such that
  \[
  s[(r_{Ai}:= b_{Ai})_i,\vec{r}_B:= s(\vec{v}_B)],h\models \Pi\land
  \bigwedge\vec{\vec{u}_B=\vec{s}_B} \land\\
  \qquad x\mapsto(\vec{t}) *
  \ast_i (\vec{R_{Ai}(v_{Ai},\vec{u}_{Ai})} \Swand
  S_{Ai}(r_{Ai},\vec{s}_{Ai})).
  \]
  Let $s' = s[(r_{Ai}:=b_{Ai})_i,\vec{r}_B:= s(\vec{v}_B)]$.
  We have $h=h_x + \Sigma_i h_{Ai}$ such that
  \[
  s',h_x\models x\mapsto(\vec{t})\\
  s',h_{Ai}\models \vec{R_{Ai}(v_{Ai},\vec{u}_{Ai})} \Swand
  S_{Ai}(r_{Ai},\vec{s}_{Ai}).
  \]
  Since $s'\models x\ne y$, we have $y\in{\rm dom}(h_{Ak})$ for some
  $k$, and hence we have
  \[
  s',h_{Ak}\models y\downarrow \land \vec{R_{Ak}(v_{Ak},\vec{u}_{Ak})} \Swand
  S_{Ak}(r_{Ak},\vec{s}_{Ak}).
  \]

  Case $s'(y)=b_{Ak}(=s'(r_{Ak}))$. Let $Q=S_{Ak}$,
  $\vec{R}_1=\vec{R}-\vec{R}_{Ak}$, and $\vec{R}_2 = \vec{R}_{Ak}$. Note
  that $Q\in\Dep(P)$ and $\vec{R}_2\subseteq\Dep(Q)$. Then, we have
  \[
  s',h_{Ak}\models \vec{R_2(v_2,\vec{u}_2)} \Swand
  Q(y,\vec{s}_{Ak}).
  \]

  We show
  \[
  s',\Sigma_{i\ne k}h_{Ai} + h_x \models
  Q(y,\vec{s}_{Ak}),\vec{R_1(v_1,\vec{u}_1)}\Swand P(x,\vec{z}).
  \]
  Now,
  \[
   \exists(r_{Ai})_i\vec{r}_B(\Pi\land x\mapsto(\vec{t}) *
  \ast_{i\ne k}S_{Ai}(r_{Ai},\vec{s}_{Ai})
  * Q(r_{Ak},\vec{s}_{Ak}) * \vec{S_B(r_B,\vec{s}_B)})
  \]
  is a definition clause of $P$.  Consider the division $\vec{R} =
  ((\vec{R}_{Ai})_{i\ne k}), (Q,\vec{R}_B)$, and it is
  sufficient to show that
\[
 s',\Sigma_{i\ne k}h_{Ai}+h_x \models \exists(r_{Ai})_{i\ne k}(
  \Pi\land \bigwedge\vec{\vec{u}_B=\vec{s}_B} \land
  \vec{s}_{Ak}=\vec{s}_{Ak} \land\\
  \qquad x\mapsto(\vec{t}) *
  \ast_{i\ne k}(\vec{R_{Ai}(v_{Ai},\vec{u}_{Ai})}\Swand
  S_{Ai}(r_{Ai},\vec{s}_{Ai})))[\vec{r}_B:=\vec{v}_B,r_{Ak}:=y],
\]
    which follows from
    \[
    s' \models \vec{\vec{u}_B=\vec{s}_B}\\
    s'(r_{Ak})=s'(y)\\
    s',h_x\models x\mapsto(\vec{t})\\
    s',h_i\models \vec{R_{Ai}(v_{Ai},\vec{u}_{Ai})}\Swand
  S_{Ai}(r_{Ai},\vec{s}_{Ai}) \qquad\mbox{(for $i\ne k$).}
    \]
    Therefore, we have
    \[
    s',h\models (Q(y,\vec{s}_{Ak}),\vec{R}_A \Swand P(x,\vec{z})) *
    (\vec{R_B(v_B,\vec{u}_B)}\Swand Q(y,\vec{s}_{Ak})),
    \]
    so we have
    \[
    s',h\models \exists\vec{w} (Q(y,\vec{w}),\vec{R}_A \Swand P(x,\vec{z})) *
    (\vec{R_B(v_B,\vec{u}_B)}\Swand Q(y,\vec{w})),
    \]
    and hence
    \[
    s,h\models \exists\vec{w} (Q(y,\vec{w}),\vec{R}_A \Swand P(x,\vec{z})) *
    (\vec{R_B(v_B,\vec{u}_B)}\Swand Q(y,\vec{w})),
    \]
    since $(r_{Ai})_i,\vec{r}_B\not\in RHS$.

    Case $s'(y)\ne b_{Ak}$.
    In this case, we have
  \[
  s',h_{Ak}\models r_{Ak}\ne y \land y \downarrow \land
  \vec{R_{Ak}(v_{Ak},\vec{u}_{Ak})} \Swand
  S_{Ak}(r_{Ak},\vec{s}_{Ak}).
  \]
  By IH, there exist $Q\in\Dep(S_{Ak})\subseteq\Dep(P)$ and a division
  $\vec{R}_{Ak}=\vec{R}_{1Ak},\vec{R}_{2Ak}$ such that
  $\vec{R}_{2Ak}\subseteq \Dep(Q)$ and
    \[
    s',h_{Ak}\models \exists\vec{w}(
    (Q(y,\vec{w}),\vec{R_{1Ak}(v_{1Ak},\vec{u}_{1Ak})}\Swand
    S_{Ak}(r_{Ak},\vec{s}_{Ak})) *
    (\vec{R_{2Ak}(v_{2Ak},\vec{u}_{2Ak})}\Swand Q(y,\vec{w}))).
    \]
    There exists $\vec{c}$ such that
    \[
    s'[\vec{w}:=\vec{c}],h_{1Ak} \models Q(y,\vec{w}),\vec{R_{1Ak}(v_{1Ak},\vec{u}_{1Ak})}\Swand
    S_{Ak}(r_{Ak},\vec{s}_{Ak})\\
    s'[\vec{w}:=\vec{c}],h_{2Ak} \models \vec{R_{2Ak}(v_{2Ak},\vec{u}_{2Ak})}\Swand Q(y,\vec{w}),
    \]
    and then we have
\[
 s'[\vec{w}:=\vec{c}],h_x + \Sigma_{i\ne k}h_i + h_{1Ak} \models
    \Pi \land \bigwedge \vec{\vec{u}_B =\vec{s}_B} \land\\
    \qquad x\mapsto(\vec{t}) *
    \ast_{i\ne k} (\vec{R_{Ai}(v_{Ai},\vec{u}_{Ai})}\Swand
    S_{Ai}(r_{Ai},\vec{s}_{Ai})) *
    (Q(y,\vec{w}),\vec{R_{1Ak}(v_{1Ak},\vec{u}_{1Ak})}\Swand
    S_{Ak}(r_{Ak},\vec{s}_{Ak})).
\]
    Let $\vec{R}_1 = (\vec{R}_{Ai})_{i\ne
    k},\vec{R}_{1Ak}$, and $\vec{R}_2 =
    \vec{R}_{2Ak}(\subseteq \Dep(Q))$, and then we have
    \[
    s'[\vec{w}:=\vec{c}],h_x + \Sigma_{i\ne k}h_i + h_{1Ak} \models
    Q(y,\vec{w}),\vec{R}_1 \Swand P
    \]
    and hence
    \[
    s'[\vec{w}:=\vec{c}],h \models
    (Q(y,\vec{w}),\vec{R}_1 \Swand P) * (\vec{R}_2\Swand Q(y,\vec{w})).
    \]
    Therefore, we have
    \[
    s,h \models
    \exists\vec{w}((Q(y,\vec{w}),\vec{R}_1 \Swand P) *
    (\vec{R}_2\Swand Q(y,\vec{w}))).
    \]
$\Box$

\begin{Lemma}[Soundness of Rule (\Factor)]\label{lemma:soundfactor}
The rule (\Factor) is $m$-sound.
\end{Lemma}

{\em Proof.}
Assume the antecedent of the conclusion is true at $(s,h)$ and
$|\Dom(h)| \le m$
in order to show the succedent of the conclusion is true at $(s,h)$.
Then the antecedent of the assumption is true at $(s,h)$.
Then the succedent of the assumption is true at $(s,h)$.
By Lemma \ref{lemma:wand1}
the succedent of the conclusion is true at $(s,h)$.
$\Box$
\section{Soundness of Rules $(\exists\ \Amalgamation1,2)$}

\begin{Lemma}\label{lemma:existsplit}
(1) 
If 
$F$ is a formula constructed from
$=,\ne,\Emp,\mapsto,\land,\lor, *, \exists x\downarrow$,
and
$T=\FV(F) \cup \{\Nil\} - \Vec y$,
and
$\Vec a \notin s(T) \cup \Dom(h) \cup \Range(h)$,
and
$\Vec a$ are different from each other,
and
\[
s,h \models \Vec y\Uparrow \land (\ne (\Vec y,T \cup \Vec y)) \land F
\]
then
\[
s[\Vec y:=\Vec a],h \models F.
\]

(2) 
If 
$\exists y\Uparrow\Phi$ is equality-full,
and $a \notin s(\FV(\exists y\Uparrow \Phi) \cup\{\Nil\}) \cup \Dom(h) \cup \Range(h)$ and
\[
s,h \models \exists y\Uparrow\Phi,
\]
then
\[
s[y:=a],h \models \Phi.
\]

(3) 
If $\exists x\Uparrow\Phi_1$ and $\exists x\Uparrow\Phi_2$ are equality-full, then
$\exists x\Uparrow\Phi_1 *
\exists x\Uparrow\Phi_2
\imp
\exists x\Uparrow(\Phi_1 * \Phi_2)$.

(4) 
If $x \in \Cells(\Phi_1)$,
and $(x \ne \FV(\exists x\Uparrow\Phi_2)) \subseteq \Phi_1$,
and $\exists x\Uparrow\Phi_2$ is equality-full,
then
$\exists x\Phi_1 *
\exists x\Uparrow\Phi_2
\imp
\exists x(\Phi_1 * (x\Uparrow \land \Phi_2))$.
\end{Lemma}

{\em Proof.}
(1)
We show the claim by induction on $F$.

If $F$ is $F_1*F_2$ or $F_1 \land F_2$ or $F_1 \lor F_2$,
the claim immediately follows from IH.

Case 1. $\exists x\downarrow F_1$,

We have $b \in \Dom(h)$ such that
\[
s[x:=b],h \models F_1.
\]
Since $\Vec a \notin \Dom(h)$, we have $\Vec a \ne b$ and $s(\Vec y) \ne b$.
From
\[
s[x:=b],h \models \Vec y\Uparrow \land (\ne (\Vec y,T \cup \{x\} \cup \Vec y)) \land F_1,
\]
by IH
\[
s[x:=b,\Vec y:=\Vec a],h \models F_1.
\]
Hence
\[
s[\Vec y:=\Vec a],h \models \exists x\downarrow F_1.
\]

Case 2. $\Emp$. The claim immediately follows.

Case 3. $t \mapsto (\Vec u)$.

$t \notin \Vec y$ since $s(t) \in \Dom(h)$.
$\Vec u \not\ni \Vec y$ since $s(\Vec u) \subseteq \Range(h)$.
Since $F$ does not contain $\Vec y$, the claim holds.

Case 4. $t \downarrow$.

$t \notin \Vec y$ since $s(t) \in \Dom(h)$.
Since $F$ does not contain $\Vec y$, the claim holds.

Case 5. $u=t$.

If $u,t \notin \Vec y$, the claim immediately follows.
If $u \equiv t$, the claim immediately follows.
If $u,t \in \Vec y$ and $u \not\equiv t$, it contradicts with $(\ne (\Vec y,T \cup \Vec y))$.
If $u \in \Vec y$ and $t \notin \Vec y$,
then $t \in T$, which contradicts with the assumption $\Vec y \ne T$.

Case 6. $u \ne t$.

If $u,t \notin \Vec y$, the claim immediately follows.
If $u \equiv t$, it contradicts with $u \ne t$.
If $u,t \in \Vec y$ and $u \not\equiv t$, 
$s[\Vec y:=\Vec a] \models u \ne t$ since $\Vec a$ are different.
If $u \in \Vec y$ and $t \notin \Vec y$,
then $t \in T$, 
so $s[\Vec y:=\Vec a] \models u \ne t$ since $\Vec a \notin s(T)$.

(2)
Let $T$ be $\FV(\exists y\Uparrow\Phi) \cup \{\Nil\}$.
Let $\Phi$ be $\exists\Vec x\exists \Vec y\Uparrow(\Pi \land \Gamma)$.
Note that $\Vec x \in \Cells(\Gamma)$.
We have $m$ such that
\[
s,h \models \exists y\Uparrow\Phi^{(m)}.
\]
We have $\Vec a \in \Dom(h)$ and $\Vec b,c$ such that
\[
s[\Vec x:=\Vec a,\Vec y:=\Vec b,y:=c],h \models \Vec yy\Uparrow \land \Pi \land \Gamma^{(m)}.
\]
Since $\Vec a \in \Dom(h)$, we have $a \ne \Vec a$.
From the equality-fullness,
\[
s[\Vec x:=\Vec a,\Vec y:=\Vec b,y:=c] \models (\ne(y\Vec y,T\cup\{\Vec x,y\Vec y\})).
\]
Choose $\Vec b'$ such that
$\Vec b' \notin s(T) \cup \Dom(h) \cup \Range(h)$,
and $\Vec b' \ne a$,
and $\Vec b'$ are different from each other.
By taking $\Vec y$ to be $y\Vec y$, $\Vec a$ to be $a\Vec b'$ 
and $s$ to be $s[\Vec x:=\Vec a,\Vec y:=\Vec b,y:=c]$ in (1),
\[
s[\Vec x:=\Vec a,\Vec y:=\Vec b', y:=a],h \models \Pi \land \Gamma^{(m)}.
\]
Hence
\[
s[y:=a],h \models \exists\Vec x\exists\Vec y\Uparrow(\Pi \land \Gamma^{(m)}).
\]
Hence
\[
s[y:=a],h \models \Phi.
\]

(3)
Assume
\[
s,h \models
\exists x\Uparrow\Phi_1 *
\exists x\Uparrow\Phi_2.
\]
We have $h_1+h_2=h$ such that
\[
s,h_1 \models \exists x\Uparrow\Phi_1, \\
s,h_2 \models \exists x\Uparrow\Phi_2.
\]
There is $a$ such that $a \notin s(\FV(\Phi_1,\Phi_2) \cup\{\Nil\}-\{x\}) \cup \Dom(h) \cup \Range(h)$.
By (2),
\[
s[x:=a],h_1 \models \Phi_1, \\
s[x:=a],h_2 \models \Phi_2.
\]
Hence
\[
s[x:=a],h_1+h_2 \models \Phi_1 * \Phi_2.
\]
Hence
\[
s,h \models \exists x\Uparrow(\Phi_1 * \Phi_2).
\]

(4)
Assume
\[
s,h \models
\exists x\Phi_1 *
\exists x\Uparrow\Phi_2.
\]
Then we have $h_1+h_2=h$ such that
\[
s,h_1 \models \exists x\Phi_1, \\
s,h_2 \models \exists x\Uparrow\Phi_2.
\]
Then we have $a \in \Dom(h_1)$,
$b$ such that
\[
s[x:=a],h_1 \models \Phi_1, \\
s[x:=b],h_2 \models x\Uparrow \land \Phi_2.
\]
Let $\Phi_2$ be $\exists \Vec x_2\exists\Vec y_2\Uparrow(\Pi_2 \land \Gamma_2)$.
Then we have $\Vec a \in \Dom(h_2)$ and $\Vec b \notin \Dom(h_2) \cup \Range(h_2)$ such that by letting $s'$ be $s[x:=b,\Vec x_2:=\Vec a,\Vec y_2:=\Vec b]$,
\[
s',h_2 \models \Pi_2 \land \Gamma_2.
\]

We will show 

Claim 1: 
$
s[x:=a],h_2 \models x\Uparrow \land \Phi_2.
$

If $a=b$, the claim immediately follows.
Assume $a \ne b$.

We can show $a \notin \Range(h_2)$ as follows.
Assume $a \in \Range(h_2)$ in order to show contradiction.
$a \ne \Vec a \Nil$ since $a \in \Dom(h_1)$.
$a \ne \Vec b$ since $\Vec b \notin \Range(h_2)$.
$a \ne s(\FV(\exists x\Uparrow\Phi_2))$ since
$(x \ne \FV(\exists x\Uparrow\Phi_2)) \subseteq \Phi_1$.
Hence $a \notin s(\FV(\exists x\Uparrow\Phi_2)) \cup \{\Vec a, \Vec b, \Nil\}$.
By Lemma \ref{lemma:leaves},
$a \in s'(\FV(\Gamma_2) \cup\{\Nil\})$, which contradicts.

We have shown $a \notin \Range(h_2)$.
Hence
$a \notin s(\FV(\exists x\Uparrow\Phi_2) \cup \{\Nil\}) \cup \Dom(h_2) \cup \Range(h_2)$.
By (2),
\[
s[x:=a],h_2 \models x\Uparrow \land \Phi_2.
\]
Hence we have shown the claim 1.

Hence
\[
s[x:=a],h \models \Phi_1 * (x\Uparrow \land \Phi_2).
\]
Hence
\[
s,h \models
\exists x(\Phi_1 * (x\Uparrow \land \Phi_2)).
\]
$\Box$

\begin{Lemma}[Soundness of Rules $(\exists\ \Amalgamation1,2)$]\label{lemma:soundamalg}
The rules $(\exists \Amalgamation1)$
and
$(\exists \Amalgamation2)$ are $m$-sound.
\end{Lemma}

{\em Proof.}
We consider both rules simultaneously.
Assume the antecedent of the conclusion is true at $(s,h)$ and
$|\Dom(h)| \le m$
in order to show the succedent of the conclusion is true at $(s,h)$.
Then the antecedent of the assumption is true at $(s,h)$.
Then the succedent of the assumption is true at $(s,h)$.
By Lemma \ref{lemma:existsplit} (3) and (4) for $(\exists\Amalgamation1)$
$(\exists\Amalgamation2)$ respectively,
the succedent of the conclusion is true at $(s,h)$.
$\Box$
\section{Soundness of Rule $(*)$}

We will show the soundness of the rule $(*)$.
It is short but one of the most interesting parts in our contribution.

\begin{Lemma}\label{lemma:distribute}
For propositional variables $A_k^i \ (k=1,2,i \in I)$,
the following is true in the propositional logic:
\[
\Land_{I_1+I_2=I} ((\Lor_{i \in I_1} A_1^i) \lor (\Lor_{i \in I_2} A_2^i))
\lequiv
\Lor_{i \in I} (A_1^i \land A_2^i).
\]
\end{Lemma}

{\em Proof.}
$\Lto:$ 
Assume the negation of the right-hand side 
\[
\neg \Lor_{i \in I} (A_1^i \land A_2^i)
\]
in order to
show the negation of the left-hand side.
It is
\[
\Land_{i \in I} (\neg A_1^i \lor \neg A_2^i)
\]
Take $I_1$ to be $\{ i \ |\ \neg A_1^i \}$ and $I_2$ be $I-I_1$.
Then
\[
(\Land_{i \in I_1} \neg A_1^i) \land (\Land_{i \in I_2} \neg A_2^i).
\]
Hence
\[
\neg ((\Lor_{i \in I_1} A_1^i) \lor (\Lor_{i \in I_2} A_2^i)).
\]
Hence the negation of the left-hand side is true.

$\Lfrom:$
Assume the $i$-th disjunct is true.
For any $I=I_1+I_2$,
$i \in I_1$ or $i \in I_2$, so the conjunct of the left-hand side is true.
$\Box$

\begin{Lemma}[Split Lemma]\label{lemma:sound-*}
The rule 
\[
\infer[(*)]{F_1 * F_2 \prove \{ G_1^i * G_2^i \ |\  i \in I \} }{
	F_1 \prove \{
G_1^i \ |\  i \in I' \}
\hbox{\ or\ }
	F_2 \prove \{ G_2^i \ |\  i \in I-I' \}
	&
	(\forall I' \subseteq I)
}
\]
is $m$-sound.
\end{Lemma}

{\em Proof.}
For each $I' \subseteq I$,
assume
\[
F_1 \models_m \{ G_1^i | i \in I' \}
\]
or
\[
F_2 \models_m \{ G_2^i | i \in I-I' \}.
\]
We will show
$
F_1 * F_2 \models_m \{ G_1^i * G_2^i | i \in I \}.
$

Assume 
$
s,h \models F_1 * F_2
$
and $|\Dom(h)| \le m$.

We have $h=h_1+h_2$ such that
$
s,h_i \models \Gamma_i \ (i=1,2).
$
Then
\[
s, h_1 \models F_1, \\
s, h_2 \models F_2.
\]
Since $|\Dom(h_1)|, |\Dom(h_2)| \le m$,
$
\Lor_{i \in I'} s,h_1 \models G_1^i
$
or
$
\Lor_{i \in I-I'} s,h_2 \models G_2^i.
$
Since this holds for each $I' \subseteq I$,
by letting $I_1 = I'$ and $I_2 = I-I'$,
\[
\Land_{I_1+I_2=I}
(\Lor_{i \in I_1} (s,h_1 \models G_1^i)
\lor
\Lor_{i \in I_2} (s,h_2 \models G_2^i)
).
\]
By Lemma \ref{lemma:distribute},
$
\Lor_{i \in I} (s,h_1 \models G_1^i \land s,h_2 \models G_2^i).
$
Therefore
$
s,h \models G_1^i * G_2^i
$
for some $i \in I$.
$\Box$

\section{Soundness of $\CSLIDomega$}

This section proves the soundness theorem of $\CSLIDomega$.
The soundness proof uses the fact that $|\Dom(h)|$ decreases upwardly by
the rule $(*\mapsto)$.

\begin{Lemma}\label{lemma:rulesound}
(1) For the rule $(*\mapsto)$,
if the assumptions are $m$-valid
then the conclusion is $(m+1)$-valid.

(2) Every rule except $(*\mapsto)$ is $m$-sound.
\end{Lemma}

{\em Proof.}
(1) Assume the antecedent of the conclusion is true at $(s,h)$ and
$|\Dom(h)| \le m+1$
in order to show the succedent of the conclusion is true at $(s,h)$.
We have $h_1+h_2=h$ such that the first conjunct is true at $(s,h_1)$
and the second conjunct is true at $(s,h_2)$.
Then the antecedent of the assumption is true at $(s,h_1)$.
Since $|\Dom(h_1)| \le m$, the succedent of the assumption is true at $(s,h_1)$.
Hence the succedent of the conclusion is true at $(s,h)$.

(2) 
Lemmas \ref{lemma:soundfactor}, \ref{lemma:soundamalg} and
\ref{lemma:sound-*} show $m$-soundness of $(\Factor)$,
$(\exists\Amalgamation)$ and $(*)$ respectively.
$(\Uparrow R)$ is $m$-sound
by Lemma \ref{lemma:existsUparrow}.
The claim for the other rules apparently holds.
$\Box$

When we allow open assumptions in a proof,
we call it an {\em open proof}.

\begin{Lemma}\label{lemma:sound}
(1) For any bud $J$ in a given proof,
$\models_m J$.

(2) If $F \prove \Vec G$ has a proof
then
$F \models_m \Vec G$ for any $m$.
\end{Lemma}

{\em Proof.}
(1) Assume $D$ is a proof.
For a bud $J$ in $D$, we define the height $|J|$ as
the number of judgments in the path from the conclusion to its companion in $D$.
Note that the path is not to the bud.
For a bud $J$ in $D$,
we will show $\models_m J$ by induction on $(m,|J|)$.

Let $J'$ be the companion of $J$.
Consider the open subproof $\pi_J$ of $J'$ in $D$.

For a bud $J_1$ in $\pi_J$ such that $|J_1| \ge |J|$,
the companion of $J_1$ is between $J'$ and $J_1$.
Hence there is a rule application $(*\mapsto)$ between $J'$ and $J_1$.
We remove the path from the assumption of the rule to $J_1$.
We do this for every such bud to obtain an open proof of $\pi_J'$ where
the open assumptions are buds of smaller heights or
the conclusions of rule applications $(*\mapsto)$.
For an open assumption $J_2$ of a bud in $\pi_J'$,
by IH for $|J_2|<|J|$, $\models_m J_2$.
Let $J_3$ be an open assumption of the conclusion of the rule $(*\mapsto)$
in $\pi_J'$.
By case analysis of $m$, we show $\models_m J_3$.

Case $m=0$.
Since the antecedent of $J_3$ is false with the empty heap,
$\models_m J_3$.

Case $m>0$.
By IH for $m-1$, the assumption of the rule $(*\mapsto)$ is $(m-1)$-valid.
By Lemma \ref{lemma:rulesound} (1), 
$J_3$ is $m$-valid.

Hence, in both cases,$\models_m J_3$.
Since $J_3$ is arbitrary,
all the open assumptions of $\pi_J'$ are $m$-valid.
By Lemma \ref{lemma:rulesound} (1)(2), $\models_m J'$ so $\models_m J$.

(2)
Assume a proof $D$ of $F \prove \Vec G$.
By (1), every bud of $D$ is $m$-valid. 
By Lemma \ref{lemma:rulesound} (1)(2), $F \models_m \Vec G$.
$\Box$

\begin{Th}[Soundness]\label{th:sound}
If $J$ is provable in $\CSLIDomega$, then $J$ is valid.
\end{Th}

{\em Proof.}
Let $J$ be $F \prove \Vec G$.
Assume $s,h \models F$.
Let $m$ be $|\Dom(h)|$.
By Lemma \ref{lemma:sound} (2), $F \models_m \Vec G$.
Hence $s,h \models \Vec G$.
$\Box$

\section{Satisfiability Checking}\label{sec:sat}

This section gives a satisfiability checking procedure
for $\psi$.

For each inductive predicate symbol $P$ of arity $m$ and each $n \ge 0$,
we add an inductive predicate symbol $\hat P^n$ of arity $m+n$.
For simplicity, we write $\hat P$ for every $\hat P^n$,
since $n$ is determined by the number of arguments.

We define the definition clauses for $\hat P(x,\Vec x,\Vec y)$ as the following (1) and (2)
for each definition clause of $P(x,\Vec x)$.
Let the definition clause of $P(x,\Vec x)$ be
\[
\exists\Vec z(\Pi \land x \mapsto (\Vec u) * *_i P_i(t_i,\Vec t_i)).
\]

(1) For each $y, (\Vec y_i)_i$ such that 
the sequence $y(\Vec y_i)_i$ is a permutation of the sequence$\Vec y$
and
the sequence $\Vec y_i$ are in the same order as the sequence $\Vec y$,
we define the definition clause of $\hat P(x,\Vec x,\Vec y)$ by
\[
\exists\Vec z(\Pi \land x = y  \land x \mapsto (\Vec u) * *_i \hat P_i(t_i,\Vec t_i,\Vec y_i)).
\]

(2) For each $(\Vec y_i)_i$ such that 
the sequence $(\Vec y_i)_i$ is a permutation of the sequence $\Vec y$ and
the sequence $\Vec y_i$ are in the same order as the sequence $\Vec y$,
we define the definition clause of $\hat P(x,\Vec x,\Vec y)$ by
\[
\exists\Vec z(\Pi  \land x \mapsto (\Vec u) * *_i \hat P_i(t_i,\Vec t_i,\Vec y_i)).
\]

\begin{Lemma}\label{lemma:satcheck-aux}
We have
$P(x,\Vec x) \land \Vec y\downarrow \lequiv \hat P(x,\Vec x,\Vec y).$
\end{Lemma}

{\em Proof.}
$\to:$

It is sufficient to show the following claim for all $m$.

Claim 1:
$P^{(m)}(x,\Vec x) \land \Vec y\downarrow \imp \hat P^{(m)}(x,\Vec x,\Vec y).$

We show this claim by induction on $m$.

Assume
\[
s,h \models P^{(m)}(x,\Vec x) \land \Vec y\downarrow
\]
in order to show $s,h \models \hat P^{(m)}(x,\Vec x,\Vec y)$.

If $m=0$, the immediately holds.
Assume $m>0$.

We have a definition clause of $P(x,\Vec x)$ 
\[
\exists\Vec z(\Pi \land x \mapsto (\Vec u) * *_i P_i(t_i,\Vec t_i))
\]
such that
\[
s,h \models \exists\Vec z(\Pi \land x \mapsto (\Vec u) * *_i P_i^{(m-1)}(t_i,\Vec t_i)).
\]

Consider cases according to $s(x) \in s(\Vec y)$.

Case $s(x) \in s(\Vec y)$.

Let $y \in \Vec y$ such that $s(x)=s(y)$.

We have $(\Vec y_i)_i$ such that 
$y(\Vec y_i)_i$ is a permutation of $\Vec y$,
$\Vec y_i$ are in the same order as $\Vec y$ and
\[
s,h \models \exists\Vec z(\Pi \land x=y \land x \mapsto (\Vec u) * *_i (P_i^{(m-1)}(t_i,\Vec t_i) \land \Vec y_i\downarrow)).
\]
By IH, 
\[
s,h \models \exists\Vec z(\Pi \land x=y \land x \mapsto (\Vec u) * *_i \hat P_i^{(m-1)}(t_i,\Vec t_i,\Vec y_i)).
\]
Hence $s,h \models \hat P^{(m)}(x,\Vec x,\Vec y)$.

Case $s(x) \notin s(\Vec y)$.

We have $(\Vec y_i)_i$ such that 
$(\Vec y_i)_i$ is a permutation of $\Vec y$,
$\Vec y_i$ are in the same order as $\Vec y$,
and
\[
s,h \models \exists\Vec z(\Pi \land x \mapsto (\Vec u) * *_i (P_i^{(m-1)}(t_i,\Vec t_i) \land \Vec y_i\downarrow)).
\]
By IH, 
\[
s,h \models \exists\Vec z(\Pi \land x \mapsto (\Vec u) * *_i \hat P_i^{(m-1)}(t_i,\Vec t_i,\Vec y_i)).
\]
Hence $s,h \models \hat P^{(m)}(x,\Vec x,\Vec y)$.

We have shown the claim 1.

$\leftarrow:$

It is sufficient to show the following claim.

Claim 2:
$\hat P^{(m)}(x,\Vec x,\Vec y) \imp P^{(m)}(x,\Vec x) \land \Vec y\downarrow.$

By induction on $m$ we show this claim.

If $m=0$, the immediately holds.
Assume $m>0$.

Assume $s,h$ satisfies the antecedent.

We have cases according to the definition clause (1) or (2) of $\hat P(x,\Vec x,\Vec y)$.

Case (1). 
Let the definition clause be
\[
\exists\Vec z(\Pi \land x = y \land x \mapsto (\Vec u) * *_i \hat P_i(t_i,\Vec t_i,\Vec y_i))
\]
and
\[
s,h \models
\exists\Vec z(\Pi \land x = y \land x \mapsto (\Vec u) * *_i \hat P_i^{(m-1)}(t_i,\Vec t_i,\Vec y_i)).
\]
By IH,
\[
s,h \models
\exists\Vec z(\Pi \land x = y \land x \mapsto (\Vec u) * *_i (P_i^{(m-1)}(t_i,\Vec t_i) \land \Vec y_i\downarrow)).
\]
Hence $s,h \models P^{(m)}(x,\Vec x)\land \Vec y\downarrow$.

Case (2) is similar to the case (1).
We have shown the claim 2.
$\Box$

We check the satisfiability of $\psi$ by extending
the decision procedure for
the satisfiability of a symbolic heap given in \cite{Brotherston14}.

\begin{Th}[\cite{Brotherston14}]
For a given symbolic heap $\phi$,
we can effectively compute the set $\Eval \phi$ of its base pairs
such that
$\phi$ is satisfiable 
iff
for some $(V,\Pi) \in \Eval \phi$,
$\Pi$ is satisfiable.
\end{Th}

\begin{Def}\label{def:satcheck}\rm
We give a decision algorithm to check the satisfiability of given
\[
Y \uparrow \land \Pi \land *_{i \in I} (X_i \downarrow \land P_i(\Vec t_i)).
\]

Step 1. By using the Lemma \ref{lemma:satcheck-aux} to eliminate $\downarrow$
keeping equivalence,
we transform the goal into 
\[
Y \uparrow \land \Pi \land *_{i \in I} \hat P_i(\Vec t_i,\Vec x_i)
\]
where $\Vec x_i$ is a sequence of elements in $X_i$.

Step 2. For each $i \in I$, we compute $\Eval{\hat P_i(\Vec t_i,\Vec x_i)}$ 
by using the algorithm by \cite{Brotherston14},
and let it be $B_i$.

Step 3. Try to find some $(V_i,\Pi_i)$ in $B_i$ for each $i \in I$ such that
$V_i$ and $Y$ are disjoint under the equality of $\Pi$,
and
$\otimes(\biguplus_{i \in I}V_i) \land \Pi \land \Land_{i \in I} \Pi_i$ is satisfiable.

Step 4. If there are such $(V_i,\Pi_i)$ $(i \in I)$, answer with "satisfiable".
Otherwise, answer with "unsatisfiable".
\end{Def}

\section{Decision Algorithm}

This section gives the algorithm to decide the validity of a given entailment.
First define normal forms and groups,
then define the algorithm,
finally we will show the partial correctness,
loop invariants, and the termination of the algorithm.

\subsection{Normal Form and Group}

This section defines a normal form and a group.
In our proof search algorithm,
a normal form is used as a bud in cyclic proofs and
the termination will be proved
by counting normal forms.
Groups are used for the $(*)$ rules to keep validity.

\begin{Def}[Groups]\rm
A {\em variable group} of $\psi$ is defined to be
the set $(\Roots+\Cells)(\Delta)$ of variables for some $\Delta$ in $\psi$.
A variable group of an entailment $J$ is defined to be a variable group of 
the antecedent in $J$.
A formula $F$ in an entailment $J$ is called a {\em group}
when $(\Roots+\Cells)(F)$ is a variable group of $J$.
In particular,
the $*$-conjunct $P(x) \land y_1 \downarrow \land \ldots \land y_n \downarrow$
in the antecedent is a group.
A formula $*_iF_i$ in an entailment $J$ is called {\em grouping} when
each $F_i$ is a group of $J$.

$(\Gamma_1,\Gamma_2)$ is called {\em group-disjoint} if
$(\Roots+\Cells)(\Gamma_1) \cap (\Roots+\Cells)(\Gamma_2) = \emptyset$.

$(\Phi_1,\Phi_2)$ is called a {\em group split} by $(\Gamma_1,\Gamma_2)$ 
if $(\Roots+\Cells)(\Gamma_i) = (\Roots+\Cells)(\Phi_i)$ for $i=1,2$.
\end{Def}
A variable group is the set of address variables that 
belong to some single $\Delta$
in the antecedent.
A group is a $*$-conjunction of formulas whose address variables are
in a single group.
$(\Gamma_1,\Gamma_2)$ is group-disjoint when
their address variables are disjoint.
$(\Phi_1,\Phi_2)$ is a group split by $(\Gamma_1,\Gamma_2)$ when
they are split by their address variables according to variable groups
of $\Gamma_1,\Gamma_2$.

For a set $V$ of variables,
we define $|P_1(t_1,\Vec t_1),\ldots, P_n(t_n,\Vec t_n)|_V$ as 
$|\{t_1,\ldots,t_n\}-V|$.

\begin{Def}[Normal Form]\label{def:normal}\rm
For a given set $V_0$ of variables and a given number $d$,
an  entailment $J$ is called {\em normal} with $(V_0,d)$ if
$J$ is of the form
$Y \uparrow \land \Pi \land \Gamma \prove
\{ \Phi_i \ |\ i \in I \}$
and $\Phi_i$ is of the form
$\exists\Vec x_i\exists \Vec y_i\Uparrow(\Pi_i \land \Gamma_i)$
and
by letting
$V$ be $\FV(J)$,
\begin{enumerate}

\item $\Gamma$ is a single group (single group condition),

\item $Y+(\Roots+\Cells)(\Gamma) = V$ (variable condition),

\item $\Roots(\Phi_i)$ is defined (disjunct root condition),

\item $(\Roots+\Cells)(\Gamma) = (\Roots+\Cells)(\Phi_i)$ for every $i \in I$ (group condition),

\item if $P(x,\Vec t) \land y\downarrow$ in $\Phi_i$, then $y \in \Vec t$
(disjunct definedness condition),

\item $\Vec x_i \subseteq \Cells(\Gamma_i)$ (disjunct existential condition), 

\item $\Vec y_i \subseteq \FV(\Gamma_i)$ (unrelated existential condition), 

\item $\Pi$ is $(\ne (V))$ (equality condition),

\item 
$\Pi_i$ is $(\ne(\Vec x_i\Vec y_i, V+\{\Vec x_i\Vec y_i,\Nil\}))$
(disjunct equality condition),

\item if $i \ne j$, then
$\Phi_i \not\equiv \Phi_j\theta$ 
for all variable renaming $\theta$ such that
$\Dom(\theta) \cap \FV(\Gamma) = \emptyset$ (disjunct renaming condition),

\item $\FV(Y,\Pi) \subseteq \FV(\Gamma,(\Phi_i)_i)$ (antecedent variable condition),

\item $|\Vec Q|_{V_0} \le d$ for every predicate symbol $\Vec Q \Swand P$ in $J$ (wand condition).

\end{enumerate}
\end{Def}

\subsection{Definition of Algorithm}

For a given entailment $A \prove \Vec B$,
our algorithm calls the function {\rm Main}($A \prove \Vec B$) to
decide whether $A \prove \Vec B$ is valid or not.
If it returns Yes,
then $A \prove \Vec B$ is valid.
If it return No,
then $A \prove \Vec B$ is not valid.
The function Main calls MainLoop,
which may fork, namely, it copies itself and produces new processes.
The function Main waits all these processes to terminate or fail.

Let $k_{\max}$ be the maximum arity for predicate symbols in
the original language.

For sets $T,T'$ of terms,
we define $(=\ne(T,T'))$ as
the set of conjunctions of all combinations of $t=u$ or $t \ne u$
for $t \in T, u \in T'$.
We define $(=\ne(T))$ as $(=\ne(T \cup \{\Nil\},T \cup \{\Nil\}))$.
We write $(\Dom \cup \Range)(f)$ for $\Dom(f) \cup \Range(f)$.

We say we apply a rule to a computation step when
the input to the step is the conclusion of the rule
and the output of the step are the assumptions of the rule.

All the functions are
defined below.
We describe functions in codes and
then describe the same functions again in English for explanation.
The algorithm of satisfiability check is given in Section \ref{sec:sat}.

\Hrule

{\small

\FunctionL{Main}{A \prove \Vec B}{

\LetS{J}{A \prove \Vec B}.

\LetS V{\FV(J)}.

\LetS{\Pi \land *_i \calP_i(x_i,\Vec t_i)}A.

$K := \{ (Y,(X_i)_i) \ |\ Y+\Sigma_i X_i+\{x_i\}_i=V \}$.

\ForeachL{(Y,(X_i)_i)}K{

$J:=(Y \uparrow \land \Pi \land *_i (\calP_i(x_i,\Vec t_i) \land X_i\downarrow) \prove \Vec B)$.

\LetS{(B_j)_{j \in J'}}{\Vec B}.

\LetS X{\Roots(A) + \Sigma_i X_i}.

$K'':= \{ ((X_{j,x})_{x \in \Roots(B_j)})_{j \in J'} \ |\ j \in J',
\Roots(B_j) + \Sigma_{x \in \Roots(B_j)} X_{j,x} = X \}$.

\ForeachL {((X_{j,x})_{x \in \Roots(B_j)})_{j \in J'}}{K''}{

\LetS{\psi \prove (\Pi_j * *_i \calP_{ji}(x_{ji},\Vec t_{ji}))_{j \in J'}}J.

$J:=(\psi \prove (\Pi_j * *_i (\calP_{ji}(x_{ji},\Vec t_{ji}) \land X_{j,x_{ji}} \downarrow))_{j \in J'})$.

$K':= (=\ne(V))$.

\ForeachL{\Pi_1}{K'}{

\LetS{\psi \prove \Vec \psi}J.

$J:=\Pi_1 \land \psi \prove \Vec \psi$.

\LetS{\psi \prove \Vec \psi}J.

\IfthenS{$\psi$ unsat}{\COntinue}

\IfthenL{$\psi$ is $V\uparrow \land \Pi \land \Emp$}{
\IfthenS{$\Pi \imp \Lor\{\Pi' \ |\ \Pi' \land \Emp \in \Vec\psi\}$ is true}{\COntinue}
\ElseS{\FAil}
}

\LetS{V}{\bigcup_{\psi_i \in \Vec\psi}\Cells(\psi_i)}.

\LetS{\Dw}{k_{\max}}.

MainLoop($\{J,V,\Dw$)
/* MainLoop will produce some processes */

Wait either of the following cases happens.

\IfthenS{some process produced by MainLoop terminates without fail}{\COntinue}

\IfthenS{every process produced by MainLoop fails}{\Return {\rm No}}

}
}
}

\Return {\rm Yes}
}

\Hrule

\FunctionL{MainLoop}{J,V,d}{

\hskip -2ex
/*
$J$ an entailment to prove,
$d$ the maximum depth of wands.
$V$ the set of cell variables.
*/

$S:=\{(J,\emptyset)\}$.

\WhileL{S \ne \emptyset}{

/*
$S$ a set of pairs of a subgoal $J$ and a history $H$.

$J$ and elements in $H$ are normal forms except the initial entailment.
*/

For each $(J,H) \in S$,
copy the current process and continue the computation

(namely, the current process becomes $|S|$ processes by fork).

$S:=S-\{(J,H)\}$.

\IfthenS{there are some $J' \in H$ and $\theta$ such that 
$J'\theta \equiv J$}{\COntinue}

$H:=H+\{J\}$

\LetS{\psi \prove \Vec\Phi}J.

\IfthenL{$\Roots(\psi) \cap \bigcap_i \Roots(\Phi_i) = \emptyset$}{
Choose $y \in \Roots(\psi)$.
/* There may be choices only for the first loop */

\LetS{\psi \prove \Vec\Phi,\Phi[P(\Vec t) \land y \downarrow]}J.

$J:={\rm Factor}(V,d, (\psi \prove \Vec\Phi,\Phi[\ ]), P(\Vec t), y)$.

}

$S':= {\rm Unfold}(J)$.

\ForeachL J{S'}{

/* Case analysis */
\LetS{\psi \prove \Vec F}{J}.
$S'' := \{ \Pi \land \psi \prove \Vec F \ |\ 
\Pi \in (=\ne(\FV(J)))
\}$.

\ForeachL{J}{S''}{

/* Unsat check */
\LetS{\psi \prove \Vec F}J.
\IfthenS{$\psi$ unsat}{\COntinue}

$J:={\rm Match}(J)$

/* In $\Emp$ case. Termination Check */

\IfthenL{$\psi \equiv Y \uparrow \land \Pi \land \Emp$}{
\IfthenS{some $\exists\Vec w\Uparrow(\Pi' \land \Emp) \in \Vec F$}{\COntinue\ /* $\Pi'$ has only $w \ne t$ */}
\FAil
}

$G:={\rm Split}(J,V,d)$

For each $S''' \in G$,
copy the current process and continue the computation

(namely, the current process becomes $|G|$ processes by fork).

\ForeachL{J}{S'''}{

$J:={\rm Normalize}(J)$

$S:=S \cup \{ (J,H) \}$.

}
}
}
}

}

\Hrule

\FunctionL{Factor}{V,d,J[\ ],P'(\Vec t),y}{

/* $J[P'(\Vec t) \land y \downarrow]$. $P'$ is in the extended language. */

/* (1) Factor */

\LetS{F \prove \Vec G,G[\ ]}{J[\ ]}.

\LetS{\Vec {Q(\Vec t)} \Swand P(\Vec t)}{P'(\Vec t)}
\WhereS{the predicate symbols $\Vec Q, P$ are in the original language.}

Apply the following rule $(\BFactor)$ to $J$.

\[
\infer[(\BFactor)]{x \ne y \land F \prove \Vec G,
G[(\Vec {Q(\Vec t)} \Swand P(x,\Vec t)) \land y\downarrow]}{
\hbox to 80ex{\vbox{\noindent $x \ne y \land F \prove \Vec G,
\{G[
\exists\Vec w(
(\Vec {Q_1(\Vec t_1)},Q(y,\Vec w) \Swand P(x,\Vec t)) *
(\Vec {Q_2(\Vec t_2)} \Swand Q(y,\Vec w))
)
]
\ |\
$
\hfil\break$
\{\Vec {Q_1(\Vec t_1)},\Vec {Q_2(\Vec t_2)}\} = \{\Vec Q(\Vec t)\},
Q \in \Dep(P), \Vec Q_2 \subseteq \Dep(Q), 
$\hfil\break$
|\Vec {Q_1(\Vec t_1)},Q(y,\Vec w)|_V \le d,|\Vec {Q_2(\Vec t_2)}|_V \le d,
\Vec w {\rm\ fresh}\}$
}}
}
\]

/* (2) Name Case Analysis */

\LetS{
\psi \prove 
F'[
\exists w\Vec w(
(\Vec {Q_1(\Vec t_1)},Q(y,\Vec u_1w\Vec u_2) \Swand P(\Vec t)) *
(\Vec {Q_2(\Vec t_2)} \Swand Q(y,\Vec u_1w\Vec u_2)))],
\Vec F
}J.

\LetS V{\FV(J)}.

\LetS{F[\ ]}{(\Vec {Q_1(\Vec t_1)},Q(y,\Vec u_1[\ ]\Vec u_2) \Swand P(\Vec t))}.

\LetS{G[\ ]}{(\Vec {Q_2(\Vec t_2)} \Swand Q(y,\Vec u_1[\ ]\Vec u_2))}.

$J:=(
\psi \prove 
F'[\exists w\Vec w(w \ne V \cup \{\Nil\} \land (F[w] \land w \downarrow) * G[w])],
\{
F'[\exists\Vec w(F[t] * G[t])]
\ |\
t \in V \cup \{\Nil\}
\},
\Vec F
)$.

Do this for each $\exists w$ produced by the previous step.

\Return J

}

\Hrule

\FunctionL{Unfold}J{

/* (1) Unfold L and R */

\LetS{\psi \prove \Vec F}J.

Choose $x \in \Roots(\psi) \cap \bigcap_i \Roots(F_i)$.

$S':=\{
\psi' * (A(x,\Vec t,\Vec z) \land X \downarrow) \prove 
\{F_i[\phi_i(x,\Vec t_i)] \ |\
$

$
F_i[\calP_i(x,\Vec t_i)] \in \Vec F,
\phi_i(x,\Vec t_i)$ is a definition clause of $P_i(x,\Vec t_i)$ if $\calP$ is $P_i$,
$\phi_i(x,\Vec t_i)$ is $\calP_i(x,\Vec t_i)$ if $\calP_i$ is $\mapsto
\}$

$\qquad
\ |\ 
\psi \equiv \psi' * (\calP(x,\Vec t) \land X \downarrow),
$

$\qquad
\exists \Vec zA(x,\Vec t,\Vec z) \hbox{\ is a definition clause of\ } P(x,\Vec t)$
if $\calP$ is $P$,
$A(x,\Vec t,\Vec z)$ is $\calP(x,\Vec t)$ if $\calP$ is $\mapsto$,
$\Vec z {\rm\ fresh}
\}
$.

/* (2) Left Definedness Distribution */

\WhileL{\hbox{$\psi * ((\Pi \land *_{i \in I} \calP_i(x_i)) \land X \downarrow) \prove \Vec F$ is in $S'$ and $|I| > 1$ and $X \ne \emptyset$}}{

$S' := S' - \{\psi * ((\Pi \land *_{i \in I} \calP_i(x_i)) \land X \downarrow) \prove \Vec F\} +
\{ \psi \land \Pi * *_i (\calP_i(x_i) \land X_i\downarrow) \prove \Vec F \ |\
\Sigma_i (X_i+\{x_i\}) = X \}$.
}

\Return{S'}

}

\Hrule

\FunctionL{Match}J{

/* (1) Equality Elimination */

\LetS{x = t \land X\uparrow \land \Pi \land \Gamma \prove \Vec F}J.

$J:=((X[x:=t]-\{\Nil\})\uparrow \land \Pi[x:=t] \land \Gamma[x:=t] 
$

\qquad$
\prove \Vec F[x:=t])$

Repeat this until the antecedent does not contain $=$.

/* (2) Match */

\LetS{\psi * x \mapsto (\Vec u) \prove 
\Vec F, \exists \Vec z(F * x \mapsto (\Vec v))}J.

$J:=(
\psi * x \mapsto (\Vec u) \prove 
\Vec F, \exists \Vec z(F * x \mapsto (\Vec v) \land \Vec u=\Vec v))$.

Do this for each $x \mapsto (\Vec v)$ in the succedent.

/* (3) Existential Instantiation */

\LetS{\psi \prove \Vec F, \exists \Vec zz(z=t \land F)}J.

$J:=
(\psi \prove \Vec F, \exists\Vec zF[z:=t])$.

Repeat this until this cannot apply anymore.

/* (4) Unmatch Disjunct Elimination */

\LetS{\psi \prove \Vec F, t=u \land F}J.
 
\IfthenS{$t \equiv u$}{$J:=(\psi \prove \Vec F, F)$}
\ElseS{$J:=(\psi \prove \Vec F)$}

Do this for each $=$ in the succedent.

/* (5) Disequality Elimination */

\LetS{\psi \prove \Vec F, t \ne u \land F}J.

\IfthenS{$t \equiv u$}{$J:=(\psi \prove \Vec F)$}
\ElseS{$J:=(\psi \prove \Vec F,F)$}

/* (6) Emp Disjunct Elimination */

\LetS{\psi \prove \Vec F, F}J.
 
\IfthenS{$F$ is $\exists\Vec x\exists\Vec y\Uparrow(\Pi \land \Emp \land X\downarrow)$ and $X\ne\emptyset$}{$J:=(\psi \prove \Vec F)$}

Do this for each disjunct.

/* (7) Unleaf Elimination */

\LetS{G \prove \Vec F, F}J.
 
\IfthenS{$F$ contains $x \mapsto (\Vec u) \land w\Uparrow$ and $w \in \Vec u$}{$J:=(G \land \Gamma \prove \Vec F)$}

Do this for each disjunct.

/* (8) $\mapsto$ Removal */

\LetS{\psi * x \mapsto (\Vec u) \prove (\exists \Vec w_i(F_i * x \mapsto (\Vec v)))_i}J.

$J:=(x \uparrow \land \psi \prove (\exists \Vec w_iF_i)_i)$.

\Return J

}

\Hrule

\FunctionL{Split}{J,V,d}{

/* (1) Extra Definedness */

\LetS{\psi \prove \Vec F,F}J.

\IfthenS{$z \in \Roots(\psi)-(\Roots+\Cells)(F)$}{$J:=(\psi \prove \Vec F,F \land z\downarrow)$.}

Repeat this until there is not $z \in \Roots(\psi)-(\Roots+\Cells)(F)$.

\WhileL{\True}{

/* (2) Right Definedness Distribution */

\LetS{\psi \prove \Vec F,
F[\exists\Vec x\exists\Vec y\Uparrow(\Pi \land F * (*_i \calP_i(x_i,\Vec t_i) \land X \downarrow))]}J.

$J:=(\psi \prove \Vec F,
\{
F[\exists\Vec x\exists\Vec y\Uparrow(\Pi \land F* *_i (\calP_i(x_i,\Vec t_i) \land X_i \downarrow))]
\ |\
X = \Sigma_i X_i
\})$.

Repeat this until the succedent becomes $\Vec\Phi$.

/* (3) Disjunct Grouping */

\IfthenL{\hbox{$J$ is $\psi \prove \Vec\Phi, \Phi[P(x,\Vec t) \land y \downarrow]$
and $x$ and $y$ are in different groups}}{
$J:={\rm Factor}(V, d,(\psi \prove \Vec\Phi,\Phi[\ ]), P(x,\Vec t),y)$.

\hskip -\Programindent
\ElseS{\BReak}
}

}

/* (4) Existential Disequality */

/* We have $\ne$ for every variable except $\Vec z$.
$\Vec x \ne \Vec z$ from the antecedent.
$\Vec y \ne \Vec z$ from $\Vec y\uparrow$ and $\Vec z\downarrow$. */

\LetS{\psi \prove \Vec\Phi,\exists\Vec x\exists\Vec y\Uparrow(\Pi \land *_i \Gamma_i)}J
\WhereS{$*_i \Gamma_i$ is grouping.}

$J:=(
\psi \prove \Vec\Phi, \exists\Vec x\exists\Vec y\Uparrow(
(\ne(\Vec x\Vec y,\FV(*_i \Phi_i) \cup \{\Nil\})) \land \Pi  \land *_i \Gamma_i) 
)$.

/* (5) Unrelatedness Introduction */

\LetS{\psi \prove \Vec\Phi,
\exists\Vec x\exists\Vec y\Uparrow(\Pi \land (\Delta \land x\downarrow) * *_i\Gamma_i)}J
\WhereS{$(\Delta \land x\downarrow) * *_i \Gamma_i$ is grouping.}

$J:=(
\psi \prove \Vec\Phi,
\exists\Vec x\exists\Vec y\Uparrow(\Pi \land (\Delta \land x\downarrow) * *_i(\Gamma_i \land x\Uparrow))
)$.

Do this for each $x\downarrow$.

/* (6) Existential Split */

\LetS{\psi \prove \Vec\Phi,\exists\Vec x\exists\Vec y\Uparrow(\Pi \land *_i \Gamma_i)}J
\WhereS{$*_i \Gamma_i$ is grouping.}

$J:=(
\psi \prove \Vec\Phi,*_i \exists\Vec x\exists\Vec y\Uparrow(\Pi \land \Gamma_i)
)$.

/* (7) $(*)$-Split */

Repeatedly apply the following rule $(\Split)$ to $J$
until the antecedent becomes a single group
in order to generate a set $G$ of sets of subgoals.
\[
\infer[(\Split)]{Y \uparrow \land \Pi \land \Gamma_1 * \Gamma_2 \prove \{ \Phi_{1i} * \Phi_{2i} | i \in I \} }{
(Y \cup Y_1) \uparrow \land \Pi \land \Gamma_2 \prove \{ \Phi_{2i} | i \in I-I' \}
{\rm\ or\ }
	(Y \cup Y_2) \uparrow \land \Pi \land \Gamma_1 \prove \{ \Phi_{1i} | i \in I' \} 
	&
	(\forall I' \subseteq I)
}
\]

where
\[
V =\FV(Y \uparrow \land \Pi \land \Gamma_1 * \Gamma_2 \prove \{ \Phi_{1i} * \Phi_{2i} | i \in I \}), \\
Y_i = (\Roots+\Cells)(\Gamma_i)  \ (i=1,2), \\
\Pi \equiv (\ne (V)), \\
(\Roots+\Cells)(\Gamma_1*\Gamma_2) + Y = V, 
\]

$(\Gamma_1,\Gamma_2)$ is group-disjoint,
and
$(\Phi_{1i},\Phi_{2i}) \ (i \in I))$ is 
group split by $(\Gamma_1, \Gamma_2)$.

\Return G

}

\Hrule

\FunctionL{Normalize}J{

/* (1) Fresh Variable Disjunct Elimination */

\LetS{X \uparrow \land \Pi \land \Gamma \prove \Vec \Phi,\Phi,\Phi'}J.

\IfthenL{$\theta$ is a variable renaming, 
$\Phi' \equiv \Phi\theta$,
$(\Dom \cup \Range)(\theta) \subseteq X$, 
$(\Dom \cup \Range)(\theta) \cap \FV(\Gamma) = \emptyset$}
{$J:=(X \uparrow \land \Pi \land \Gamma \prove \Vec \Phi,\Phi)$}

Repeat this until it cannot apply.

/* (2) Unnecessary Disequality Elimination */

\LetS{X \uparrow \land x \ne t \land \Pi \land \Gamma \prove \Vec\Phi}J.

\IfthenS{$x \notin \FV(\Gamma,\Vec\Phi)$}
{$J:=(X \uparrow \land \Pi \land \Gamma \prove \Vec\Phi)$}

Repeat this until it cannot apply.

/* (3) Unnecessary Variable Elimination */

\LetS{x \uparrow \land X \uparrow \land \Pi \land \Gamma \prove \Vec\Phi}J.

\IfthenS{$x \notin \FV(\Gamma,\Vec\Phi)$}{
$J:=(X \uparrow \land \Pi \land \Gamma \prove \Vec\Phi)$}

Repeat this until it cannot apply.

/* (4) Unnecessary Unrelated Existential Elimination */

\LetS{\psi\prove \Vec\Phi,\exists\Vec x\exists y\Vec y\Uparrow(\Pi \land \Gamma)}J.

\IfthenL{$y \notin \FV(\Gamma)$}{
Remove $y \ne t$ in $\Pi$ to obtain $\Pi'$.

$J:=(\psi \prove \Vec\Phi,\exists\Vec x\exists\Vec y\Uparrow(\Pi' \land \Gamma))$
}

Repeat this until it cannot apply.

\Return J
}

}

\Hrule

For explanations of these functions
we give their English descriptions except {\rm Mainloop} below.

\Hrule

\FunctionL{Main}{A \prove \Vec B}{

1. 
In the antecedent $A$,
consider cases
by classifying all the free variables except the roots 
into
$Y \uparrow$ and $X_i \downarrow$ for each atomic spatial formula
$P_i(x_i)$,
and in each case,
add $Y \uparrow\land \ $
and $X_i \downarrow\land\ $ to the $*$-conjunct $P_i(x_i)$,
and do the following.

2. 
In each disjunct of the succedent $\Vec B$,
consider cases
by classifying the address variables of the antecedent
into $X_i$ and $x_i$ for each atomic spatial formula $Q_i(x_i)$,
and in each case,
generate a new disjunct by
adding $X_i \downarrow\land\ $ to the $*$-conjunct $Q_i(x_i)$,
and do the following.

3. Consider cases
by $=$ and $\ne$ for all the variables and $\Nil$,
and in each case,
add $\Pi_1 \land\ $ to the antecedent where $\Pi_1$ is the conjunction
of these $=$ and $\ne$,
and do the following.

4. If the antecedent is unsatisfiable, finish this case.

5. If the spatial part of the antecedent is $\Emp$, 
do the following.
If there is some disjunction whose spatial part is $\Emp$ and
pure part is included in the pure part of the antecedent,
finish this case.
Otherwise stop with fail.

6. 
Let $J$ be the entailment.
Let $V$ be the union of $\Cells$ for each disjunct in the succedent.
Let $\Dw$ be $k_{\max}$.
Call {\rm MainLoop($\{J,V,\Dw$)}.
MainLoop may fork and process several processes.
Wait either of the following cases happens:
If some process terminates without fail, finish this case;
If every process fails, stop with No.

7. If all the cases are finished, stop with Yes.

}

\Hrule

\FunctionL{Factor}{V,d,J[\ ],P'(\Vec t),y}{

1. Factor.
Let $J[\ ]$ be $F \prove \Vec G,G[\ ]$.
Let $P'(\Vec t)$ be
$\Vec {Q(\Vec t)} \Swand P(\Vec t)$
where the predicate symbols $\Vec Q, P$ are in the original language.
Apply the following rule $(\BFactor)$ to $J$.

\[
\infer[(\BFactor)]{x \ne y \land F \prove \Vec G,
G[(\Vec {Q(\Vec t)} \Swand P(x,\Vec t)) \land y\downarrow]}{
\hbox to 80ex{\vbox{\noindent $x \ne y \land F \prove \Vec G,
\{G[
\exists\Vec w(
(\Vec {Q_1(\Vec t_1)},Q(y,\Vec w) \Swand P(x,\Vec t)) *
(\Vec {Q_2(\Vec t_2)} \Swand Q(y,\Vec w))
)
]
\ |\
$
\hfil\break$
\{\Vec {Q_1(\Vec t_1)},\Vec {Q_2(\Vec t_2)}\} = \{\Vec Q(\Vec t)\},
Q \in \Dep(P), \Vec Q_2 \subseteq \Dep(Q), 
$\hfil\break$
|\Vec {Q_1(\Vec t_1)},Q(y,\Vec w)|_V \le d,|\Vec {Q_2(\Vec t_2)}|_V \le d,
\Vec w {\rm\ fresh}\}$
}}
}
\]

2. Name Case Analysis.
For each $\exists w$ generated in the step 1, do the following.
Let $V_1$ be the union of the set of free variables in $J$
and the set of bound variables whose scopes include this $\exists w$.
Generate disjuncts by classifying cases
by $=$ and $\ne$ between
each element in $V_1 \cup \{\Nil\}$ and $w$.
If a disjunct has $w=t$, eliminate $\exists w$ by substituting $w:=t$
in the disjunct.
If a disjunct has $w \ne V_1 \cup\{\Nil\}$,
add $w\downarrow\land\ $ to 
$\Vec {Q_1(\Vec t_1)},Q(y,\Vec w) \Swand P(\Vec t)$
in the disjunct.

3. Return the entailment.

}

\Hrule

\FunctionL{Unfold}J{

1. Unfold L and R.
In the entailment $J$,
choose a common root $x$ among the antecedent and all disjuncts of the succedent.

Generate a set of subgoals by
replacing $P(x)$ in the antecedent by formulas
obtained from the definition clauses of $P(x)$ by removing $\exists \Vec z$
and replacing $\Vec z$ by fresh variables, if $P(x)$ is in the antecedent.
In each entailment in the subgoal set,
replace $Q(x)$ by each definition clause of $Q(x)$ to generate new disjuncts,
if $Q(x)$ in the disjunct.

2. Left Definedness Distribution.
In the antecedent of each subgoal entailment $J$ in the subgoal set,
consider cases of
distributing $\ \land x\downarrow$ 
to spatial atomic formulas,
and replace $J$ by 
new subgoals generated from these cases.

3. Return the subgoal set.

}

\Hrule

\FunctionL{Match}J{

1. Equality Elimination.
In the entailment $J$, if the antecedent has $x=t$,
eliminate it by substitution $x:=t$.

2. Match.
If the antecedent has $x \mapsto (\Vec u)$
and a disjunct in the succedent has $x \mapsto (\Vec v)$,
add $\Vec u=\Vec v\land\ $ to $x \mapsto (\Vec v)$.

3. Existential Instantiation.
If the succedent has $\exists z(z=t \land \ldots)$,
eliminate $\exists z$ by substitution $z:=t$.

4. Unmatch Disjunct Elimination.
If the succedent has $t=t$, remove it.
If a disjunct of the succedent has $t=u$ and $t \not\equiv u$,
remove the disjunct.

5. Disequality Elimination.
If a disjunct of the succedent has $t \ne t$, remove the disjunct.
If the succedent has $t \ne u$ and $t \not\equiv u$,
remove $t \ne u$.

6. Emp Disjunct Elimination.
If a disjunct of the succedent has $\Emp \land t\downarrow$, remove this disjunct.

7. Unleaf Elimination.
If a disjunct of the succedent has $x \mapsto (\Vec u) \land u_i\Uparrow$
and $u_i \in \Vec u$, remove the disjunct.

8. $\mapsto$ Removal.
If the antecedent and all the disjuncts of the succedent have $x \mapsto (\Underscore)$,
remove these $x \mapsto (\Underscore)$ and add
$x\uparrow\land\ $ to the antecedent.

9. Return the entailment.
}

\Hrule

\FunctionL{Split}{J,V,d}{

1. Extra Definedness.
For each root $z$ of the antecedent and each disjunct $F$ of the succedent,
if $z \notin (\Roots+\Cells)(F)$, add $z\downarrow \land$ to $F$.
Repeat this until there are not such $z$ and $F$.

2. Right Definedness Distribution.
In each disjunct in the succedent of the entailment $J$,
consider cases of
distributing $\ \land x\downarrow$ 
to spatial atomic formulas,
and replace the original disjunct by 
all the disjuncts generated from these cases.

3. Disjunct Grouping.
  If some disjunct has $P(x,\Vec t) \land y\downarrow$ and $x$ and $y$ are
in different groups, 
by letting the entailment be $\psi \prove \Vec\Phi,\Phi[P(x,\Vec t)]$,
call ${\rm Factor}(V, d,(\psi \prove \Vec\Phi,\Phi[\ ]), P(x,\Vec t),y)$.
Then go to the step 2.

4. Existential Disequality.
In each disjunct,
add $w \ne z \land$ to the $\exists$-body for each $w,z$ such that
the disjunct has $\exists w$ and $z$ is a fresh variable introduced by 
the function Unfold.

5. Unrelatedness Introduction.
Replace
each disjunct 
$\exists\Vec x(\Pi \land (\Phi \land x\downarrow) * *_i\Phi_i \land Y\Uparrow)$
by
$\exists\Vec x(\Pi \land (\Phi \land x\downarrow) * *_i(\Phi_i \land x\Uparrow))\land Y\Uparrow$,
where
$(\Phi \land x\downarrow) * *_i\Phi_i$ is grouping.

6. Existential Split.
Replace each disjunct $\exists\Vec x(\Pi \land *_i \Phi_i \land Y\Uparrow)$
by
$*_i \exists\Vec x(\Pi \land \Phi_i \land Y\Uparrow)$.

7. $(*)$-Split.
Apply the following rule $(\Split)$ repeatedly 
to generate the subgoal set $G$ until 
the spatial part of the antecedent in every entailment becomes atomic.
\[
\infer[(\Split)]{Y \uparrow \land \Pi \land \Gamma_1 * \Gamma_2 \prove \{ \Phi_{1i} * \Phi_{2i} | i \in I \} }{
(Y \cup Y_1) \uparrow \land \Pi \land \Gamma_2 \prove \{ \Phi_{2i} | i \in I-I' \}
{\rm\ or\ }
	(Y \cup Y_2) \uparrow \land \Pi \land \Gamma_1 \prove \{ \Phi_{1i} | i \in I' \} 
	&
	(\forall I' \subseteq I)
}
\]
where
\[
V =\FV(Y \uparrow \land \Pi \land \Gamma_1 * \Gamma_2 \prove \{ \Phi_{1i} * \Phi_{2i} | i \in I \}), \\
Y_i = (\Roots+\Cells)(\Gamma_i)  \ (i=1,2), \\
\Pi \equiv (\ne (V)), \\
(\Roots+\Cells)(\Gamma_1*\Gamma_2) + Y = V, 
\]
$(\Gamma_1,\Gamma_2)$ is group-disjoint,
and
$(\Phi_{1i},\Phi_{2i}) \ (i \in I))$ is 
group split by $(\Gamma_1, \Gamma_2)$.

8. Return $G$.

}

\Hrule

\FunctionL{Normalize}J{

1. Fresh Variable Disjunct Elimination.
If the entailment $J$ has the form
$X \uparrow \land \Pi \land \Gamma \prove \Vec \Phi,\Phi,\Phi'$
and there is a variable renaming $\theta$ such that
$\Phi' \equiv \Phi\theta$,
$(\Dom \cup \Range)(\theta) \subseteq X$, 
and
$(\Dom \cup \Range)(\theta) \cap \FV(\Gamma) = \emptyset$,
then
remove $\Phi'$.

2. Unnecessary Disequality Elimination.
If $x$ does not occur in 
neither the spatial part and $\downarrow$ of the antecedent or the succedent,
then remove $x \ne t$ in the antecedent.

3. Unnecessary Variable Elimination.
If $x$ does not occur in 
neither the spatial part and $\downarrow$ of the antecedent or the succedent,
then remove $x \uparrow$ in the antecedent.

4. Unnecessary Unrelated Existential Elimination.
If $y$ does not occur in the spatial part in a disjunct with $\exists y\Uparrow$, 
remove $\exists y\Uparrow$ and $y \ne t$ in the disjunct.

5. Return the entailment.

}

\Hrule

When an entailment $A \prove \Vec B$ is given to {\rm Main},
we generate subgoals 
by putting $\downarrow$ and $\uparrow$ of all the free variables 
to the antecedent with all the cases,
putting $\downarrow$ of all the address variables in the antecedent to
the succedent with all the cases,
putting $=$ and $\ne$ of all the variables with \Nil\ to the antecedent
with all cases,
check satisfiability,
and send the subgoal set to the function MainLoop.
MainLoop may fork and produce several processes.
If for some subgoal every process of MainLoop fails, the function Main returns No.
If for all subgoals some process of MainLoop terminates without fail, 
the function Main returns Yes.

{\rm MainLoop} first checks whether the current subgoal appeared already.
If so, we finish this subgoal and go to the next subgoal.
Secondly {\rm MainLoop} calls {\rm Factor} to create common roots
on both sides when there is no common root.
In order to find a common root such that
the bounded factor rule for it is locally complete,
the function MainLoop forks and produces new processes with each common root.
When there is a common root $x$, 
{\rm MainLoop} first unfolds 
the predicate of form $P(x)$ on both sides (the function {\rm Unfold}).
Next
it forces match of them on both sides by putting equalities on the succedent
and removes $x \mapsto (\Underscore)$ on both sides
(the function {\rm Match}).
Then {\rm MainLoop} checks termination condition when the antecedent is $\Emp$.
Then {\rm MainLoop} splits $\exists$-scopes.
Then we split separating conjunctions until
every entailment become a single group (the function {\rm Split}).
For each rule instance of $(\Split)$,
{\rm MainLoop} forks and produces new processes to compute this case,
and executes these processes nondeterministically.
Then {\rm MainLoop} transforms subgoals into normal forms and
goes to the first step of the function MainLoop.
The function MainLoop does this loop until the subgoal set becomes empty.

\subsection{Partial Correctness of Algorithm}

We can show the correctness of the algorithm when it terminates with Yes.
We will show the case with No in the completeness proof later.

\begin{Lemma}[Partial Correctness]\label{lemma:partialcorrect}
If the algorithm terminates with Yes, then the input entailment is provable.
\end{Lemma}

{\em Proof.}
When the algorithm terminates with Yes, 
for each subgoal there is some process that
terminates without fail.

For the input goal, the function Main generates subgoals for it.
For each subgoal, the function MainLoop constructs its proof
when it terminates without fail.
In the while loop of the function MainLoop,
for each subgoal $(J,H) \in S$,
the function MainLoop
proves $J$ by using $H$ as companions.

We can show that
each step of the algorithm consists of applications of inference rules
as follows.
We consider each step of each function.

Function {\rm Main}.

Case analysis by $(Y,(X_i)_i)$.
By the rules $(\Case\ L)$ and $(\downarrow\Out\ L)$.

Case analysis by $K''$. By $(\land\Elim)$ and $(\downarrow\ \Out)$.

Case analysis by $K'$: By rule $(\Case\ L)$.

Unsatisfiability check. By $(\Unsat)$.

Termination check: By $(\Emp)$, $(\land L)$, $(\land R)$, $(\lor R)$.

By MainLoop, each subgoal $J$ is provable.

Function {\rm MainLoop}.

If MainLoop terminates without fail for  the input $\{(J,H)\}$,
$J$ is provable by using $H$ as companions.

We use rule $(\Subst)$ from $J'$ to $J$ 
and discharge $J$ as a bud with the companion $J' \in H$.
Since $H$ is the set of judgments from the root to the previous subgoal,
the step for $J' \in H$ and $J'\theta = J$ in {\rm MainLoop} gives a bud-companion relation.

Case Analysis: By rules $(\Case\ L)$.

Unsat Check: By rule $(\Unsat)$.

Termination Check: 
By $(\Emp)$, $(\Uparrow R)$, $(\land L)$, $(\lor R)$.

Function {\rm Factor}. We show the input formula is derived from the output formula.

(1) Factor. By Lemma \ref{lemma:wand1}.

(2) Name Case Analysis.
By the rules $(\land\Elim)$ and $(\exists R)$.

Function {\rm Unfold}. 
We show that $J$ is derivable from $S'$.

(1) Unfold L and R. By the rules $(\Pred\ L)$ and $(\Pred\ R)$.

(2) Left Definedness Distribution. By $(\downarrow\Out\ L)$.

Function {\rm Match}. We show the input formula is derived from the output formula.

(1) Equality Elimination. By the rule $(=L)$.

(2) Match. By the rule $(\land\Elim)$.

(3) Existential Instantiation. By the rules $(=R)$ and $(\exists R)$.

(4) Unmatch Disjunct Elimination. By the rules $(=R)$ and $(\lor R)$.

(5) Disequality Elimination. By $(\land R)$, $(\lor R)$.

(6) Emp Disjunct Elimination. By the rule $(\lor R)$.

(7) Unleaf Elimination. By the rule $(\lor R)$.

(8) $\mapsto$ Removal. By the rule $(*\mapsto)$.

Function {\rm Split}. We show that $J$ is derived from any subgoal set in $S$.

(1) Extra Definedness. By $(\downarrow\ R)$.

(2) Right Definedness Distribution. By the rule $(\downarrow\ \Out\ R)$.

(3) Disjunct Grouping. By Factor.

(4) Existential Disequality. By the rule $(\land\Elim)$.

(5) Unrelatedness Introduction. By the rule $(\land\Elim)$.

(6) Existential Split. 
Use $(\exists\Amalgamation2)$ for $\Vec x$ and
$(\exists\Amalgamation1)$ for $\Vec y\Uparrow$.

(7) $(*)$-Split. The rule $(\Split$) is derivable from the rule $(*)$
and $(\uparrow\Elim)$.

Function {\rm Normalization}. We show that the input formula is derived from the output formula.

(1) Fresh Variable Disjunct Elimination. By the rule $(\lor R)$.

(2) Unnecessary Disequality Elimination. By the rule $(\land L)$.

(3) Unnecessary Variable Elimination. By the rule $(\lor L)$.

(4) Unnecessary Unrelated Existential Elimination. By $(\Uparrow R)$.

We have shown that
each step of the algorithm consists of applications of inference rules.
Hence the process produces a proof of the input entailment,
when it terminates without fail.
Hence the function Main produces the proof when it returns Yes.
$\Box$

\subsection{Loop Invariant}

We will use the following loop invariants of the function MainLoop.

\begin{Lemma}\label{lemma:invariant}
Let $V,\Dw$ be those in the call of the function {\rm MainLoop}.
At the beginning of the while loop in the function {\rm MainLoop},
every entailment $(J,H) \in S$ satisfies the following.

\begin{enumerate}

\item $J$ is a normal form with $(V,\Dw)$ except
the single group condition,
the disjunct equality condition, and
the disjunct renaming condition.

\item $J$ is a normal form with $(V,\Dw)$ 
after the first loop.

\item the antecedent of $J$ is satisfiable.

\item $\Cells(\Gamma) \subseteq \Cells(\Gamma')$ where
$\Gamma$ is the spatial part of antecedent in $J$ and
$\Gamma'$ is that in the initial goal given to the function MainLoop.

\end{enumerate}

\end{Lemma}

{\em Proof.}

1 and 2.

The single group condition. By the function Split.

The variable condition. This holds for the initial entailment.
New variables are $\Vec z$ in the function Unfold, and they are in the roots.

The disjunct root condition is trivial.

The group condition.
$J$ sent to MainLoop satisfies this condition.
New variables $\Vec z$ introduced by Unfold is added to the succedent
by the extra definedness step in Split.

The disjunct definedness condition.
By the name case analysis step of Factor.

The disjunct existential condition.
By the name case analysis step of Factor.

The unrelated existential condition.
By the unrelatedness introduction step of Split.

The equality condition. By case analysis in the function Main.

The disjunct equality condition. 
By the existential disequality step of Split.

The disjunct renaming condition
and the antecedent variable condition.
By the function Normalize.

The wand condition. By the function Factor.

3. By the unsatisfiability check in the functions Main and MainLoop.

4. The algorithm does not increase $\Cells(\Gamma)$.
$\Box$

\subsection{Termination}

This subsection shows the termination of the algorithm.
First we show the finiteness of the set of normal forms
possibly used in the algorithm.
Then we show the termination by using the finiteness.

Since 
a normal form during the loop consists of a single group,
the number of normal forms up to variable renaming is proved to be finite.

\begin{Lemma}\label{lemma:finite}
The set of normal forms with $(V_0,d)$ up to variable renaming is finite.
\end{Lemma}

{\em Proof.}

$d$ is the maximum depth of wands.
Let $J$ be the entailment sent to {\rm MainLoop} from {\rm Main},
$V$ be $\FV(J)$,
$Y\uparrow \land \Pi \land \Gamma$ be the antecedent of $J$,
$(\Phi_i)_{i \in I}$ be the succedent of $J$,
$c_1$ be $|\Cells(\Gamma)|$,
$k_{\max}$ be the maximum arity of
original inductive predicates and the $\mapsto$ predicate,
$c_4$ be the number of
original inductive predicates and the $\mapsto$ predicate.

We will count numbers of normal forms up to variable renaming.
We count $\Uparrow$ as a single predicate symbol.

First
the maximum number $k_1$ of arguments for 
extended inductive predicates and the $\mapsto$ predicate
is $\le k_{\max}(d+1)$.

The number $k_2$ of
extended inductive predicates and the $\mapsto$ predicate
is $\le c_4^{d+1}$.

The number $k_3$ of $*$-conjuncts in $\Gamma$ 
is $1$ by the single group condition.

The number $k_4$ of arguments in $\Gamma$ is
$\le k_3k_1+c_1$,
since 
the number of arguments for inductive or $\mapsto$ predicates is $\le k_3k_1$
and
the number of arguments for $\downarrow$ is $\le c_1$.

The number $k_5$ of variables in $\Gamma$ is $\le k_4$.

The number $k_6$ of $*$-conjuncts in $\Phi_i$ 
is $\le k_3+c_1$, since 
$P(x)$ in $\Phi_i$ implies $x \in (\Roots+\Cells)(\Gamma)$.

The number $k_7$ of arguments for inductive predicates and $\mapsto$ in  $\Phi_i$ is
$\le k_1k_6$, since the number of arguments for inductive or $\mapsto$ predicates is $\le k_1$ and the number of the predicate symbols is $\le k_6$.

The number $k_8$ of variables in $\exists$-body of $\Phi_i$ is
$\le k_7$.

The number $k_9$ of disjuncts in the antecedent (namely $|I|$) is
$\le k_2^{k_6}(k_8+1)^{k_7}2^{2k_7}$
since
the number of combination for inductive predicates and $\mapsto$ is $\le k_2^{k_6}$,
the number of combination for their arguments is $\le (k_8+1)^{k_7}$
(note that +1 for $\Nil$),
and the number of choice for existentials is $\le 2^{k_7}$
for each of $\Vec x$ and $\Vec y\Uparrow$.

The number $k_{10}$ of $\Gamma$ is
$\le k_{2}^{k_3}(k_5+1)^{k_4}$,
since
the number of combination for predicates is $\le k_{2}^{k_3}$,
and
the number of combination for arguments is $\le (k_5+1)^{k_4}$.
(note that +1 for $\Nil$.)

The number $k_{11}$ of $\Gamma \prove (\Phi)_{i \in I}$ is
$\le k_{10}2^{k_9}$,
since 
$Y\uparrow$ and $\Pi$ are determined uniquely by $\Gamma$,
the number of the antecedent is $\le k_{10}$ and
the number of succedents is $\le 2^{k_9}$.
$\Box$

\begin{Lemma}[Termination]\label{lemma:terminate}
(1) Every process produced by the function MainLoop terminates or fails.

(2) The algorithm always terminates.
\end{Lemma}

{\em Proof.}
(1) Assume some process does not terminate or fail in order to show contradiction.
By Lemma \ref{lemma:finite}, we have some upper bound $n$ for
the number of normal forms.
The step $H:=H+\{J\}$ in {\rm MainLoop} increases $|H|$.
After the $(n+2)$-th execution of the while loop in {\rm MainLoop} of the process,
$|H|=n+2$. Since 
every element in $H$ except one element is a normal form
and they are not equal by variable renaming, it contradicts.

(2) By (1).
$\Box$

\begin{Prop}
The algorithm is nondeterministic double exponential time.
\end{Prop}

{\em Proof.}
By the upper bound given by the proof of Lemma \ref{lemma:finite}.
$\Box$

\section{Constant Store Validity}

\begin{Def}\rm
For a bijection $\beta:\Locs \to \Locs$,
we define
\[
\beta(s) = \beta \circ s, \\
\beta(h) = \beta \circ h \circ \beta^{-1}.
\]
\end{Def}

\begin{Lemma}\label{lemma:heaptransformation}
For a formula $F$ in the extended language,
\[
s,h \models F
\]
and $\beta:\Locs \to \Locs$ is a bijection, then
\[
\beta(s),\beta(h) \models F.
\]
\end{Lemma}

{\em Proof.}

First we show:

Claim 1: the claim holds when $F$ does not contain inductive predicates.

We can show the claim 1 by induction on $F$.
Every case is straightforward.
We show only the case $\neg G$.

Case $\neg G$.

Assume $s,h \models \neg G$ in order to show $\beta(s),\beta(h) \models \neg G$.
Then $s,h \not\models G$.
Hence $\beta^{-1}(\beta(s)),\beta^{-1}(\beta(h)) \not\models G$.
By IH, $\beta(s),\beta(h) \not\models G$.
Hence $\beta(s),\beta(h) \models \neg G$.

We have proved the claim 1.

Next we show the claim of the lemma by induction on $F$.
We show only the case of an inductive predicate.
The other cases are proved straightforwardly.

Case $P(\Vec t)$.

Assume $s,h \models P(\Vec t)$.
We have $m$ such that $s,h \models P^{(m)}(\Vec t)$.
By the claim (1), $\beta(s),\beta(h) \models P^{(m)}(\Vec t)$.
Hence $\beta(s),\beta(h) \models P(\Vec t)$.
$\Box$

\begin{Def}\rm
$A \models_s \{ B_i | i \in I \}$ is defined by
\[
\forall h(s,h \models A \imp \Lor_{i \in I} s,h \models B_i).
\]
\end{Def}

\begin{Lemma}\label{lemma:constantstore}
Let $F_i, G_i^j$ be formulas.
If
\[
\Pi \supseteq 
(\ne \FV(\Pi,(F_j,G_j^i)_{i \in I_j, j=1,2}) ), \\
\forall s(
(\Pi \land F_1 \models_s \{ G_1^i | i \in I_1 \}) \lor
(\Pi \land F_2 \models_s \{ G_2^i | i \in I_2 \})
)
\]
then
\[
(\Pi \land F_1 \models \{ G_1^i | i \in I_1 \}) \lor
(\Pi \land F_2 \models \{ G_2^i | i \in I_2 \}).
\]
\end{Lemma}

{\em Proof.}
Let $\Vec y = \FV((F_j,G_j^i)_{i \in I_j, j=1,2})$.

Assume 
\[
(\Pi \land F_1 \models \{ G_1^i | i \in I_1 \}) \lor
(\Pi \land F_2 \models \{ G_2^i | i \in I_2 \})
\]
does not hold.
Then we have $s_1,h_1,s_2,h_2$ such that
\[
s_1,h_1 \models \Pi \land F_1, \\
s_1,h_1 \not\models \{ G_1^i | i \in I_1 \}, \\
s_2,h_2 \models \Pi \land F_2, \\
s_2,h_2 \not\models \{ G_2^i | i \in I_2 \}.
\]
Let $s_1(\Vec y) = \Vec a$ and $s_2(\Vec y) = \Vec b$.
Since $s_1 \models \Pi$ and $s_2 \models \Pi$,
we have $\Vec a \ne \Nil$ and $\Vec b \ne \Nil$,
$\Vec a$ are in $\Locs$ and distinct to each other, and
$\Vec b$ are in $\Locs$ and distinct to each other.
Take some bijection $\beta:\Locs \to \Locs$
such that $\beta(a_i)=b_i$.
Then $s_2=_{\Vec y} \beta(s_1)$.
By Lemma \ref{lemma:heaptransformation},
\[
\beta^{-1}(s_2),\beta^{-1}(h_2) \models \Pi \land F_2, \\
\beta^{-1}(s_2),\beta^{-1}(h_2) \not\models \{ G_2^i | i \in I_2 \}.
\]
Hence
\[
s_1,\beta^{-1}(h_2) \models \Pi \land F_2, \\
s_1,\beta^{-1}(h_2) \not\models \{ G_2^i | i \in I_2 \}.
\]
By taking $s$ to be $s_1$ in the assumption,
\[
(\Pi \land F_1 \models_{s_1} \{ G_1^i | i \in I_1 \}) \lor
(\Pi \land F_2 \models_{s_1} \{ G_2^i | i \in I_2 \})
\]

Case 1. $\Pi \land F_1 \models_{s_1} \{ G_1^i | i \in I_1 \}$.
Since 
\[
s_1,h_1 \models \Pi \land F_1, \\
s_1,h_1 \not\models \{ G_1^i | i \in I_1 \}
\]
we have contradiction.

Case 2. $\Pi \land F_2 \models_{s_1} \{ G_2^i | i \in I_2 \}$.
Since
\[
s_1,\beta^{-1}(h_2) \models \Pi \land F_2, \\
s_1,\beta^{-1}(h_2) \not\models \{ G_2^i | i \in I_2 \}.
\]
we have contradiction.

Since every case leads to contradiction,
we have
\[
(\Pi \land F_1 \models \{ G_1^i | i \in I_1 \}) \lor
(\Pi \land F_2 \models \{ G_2^i | i \in I_2 \})
\]
$\Box$

\section{Cone}

\begin{Def}\rm
For a heap $h$ and $a \in \Val$,
the heap $h \Downarrow a$ is defined as
$h|_X$ where $X$ is the least fixed point of
\[
F(X) = (\{ a \} \cap \Dom(h)) \cup \{ c \in \Dom(h) | b \in X, b \To c \}.
\]
$b \To c$ denotes $(h(b))_k=c$ for some $k$.
We call $h \Downarrow a$ a {\em cone} of root $a$ in the heap $h$.
For $b \in h \Downarrow a$, the depth of $b$ in $h \Downarrow a$
is defined as the least number $d$ such that $F^{d+1}(\emptyset) \ni b$.
We write $b \to c$ when $b \To c$ and $c \in \Dom(h)$.
We write $\toto$ to the reflexive and transitive closure of $\to$.
We write $b \to_\Tr c \in h\Downarrow a$ when 
$b \to c$,
$F^d(\emptyset) \ni b$ and
$F^d(\emptyset) \not\ni c$ for some $d$.
We write $b \to_\Bk c \in h \Downarrow a$ 
when $b \to c$ holds and $b \to_\Tr c \in h\Downarrow a$ does not hold.

For a heap $h$, $a \in \Val$ and $S \subseteq \Val$,
the heap $h \Downarrow_S a$ is defined as
$(h |_{\Dom(h)-(S - \{a\})}) \Downarrow a$,
and we call it a cone of root $a$ with guard $S$ in the heap $h$.
\end{Def}

We write $a \in h$ for $a \in \Dom(h)$, and
$h \subseteq h'$ for $h = h'|_{\Dom(h)}$.

\begin{Lemma}\label{lemma:cone}
(1) If
\[
s,h \models \Gamma,
\]
then
\[
h = \bigcup_{x \in \Roots(\Gamma)}h \Downarrow s(x).
\]

(2) If $h=h_1+h_2$ and $\Gamma$ is $\Gamma_1*\Gamma_2$ and for $i=1,2$
\[
s,h_i \models \Gamma_i \land (\FV(\Gamma)-(\Roots+\Cells)(\Gamma_i))\uparrow,
\]
then
\[
h_i = \bigcup_{x \in (\Roots+\Cells)(\Gamma_i)} h \Downarrow_{s((\Roots+\Cells)(\Gamma))} s(x).
\]
\end{Lemma}

{\em Proof.}
(1) 
Let $\Gamma$ be $*_i(\calP_i(x_i,\Vec t_i) \land X_i\downarrow)$.
Then we have $h_i$ such that $h=\Sigma_ih_i$ and
\[
s',h_i \models \calP_i(x_i,\Vec t_i) \land X_i\downarrow.
\]
Hence
\[
h_i \subseteq h \Downarrow s(x_i).
\]
Hence
\[
h \subseteq \bigcup_i(h \Downarrow s(x_i)).
\]
Hence
\[
h = \bigcup_{x \in \Roots(\Gamma)}h \Downarrow s(x).
\]

(2)
We write $RC$ for $(\Roots+\Cells)$ for simplicity.

By (1),
\[
h = \bigcup_{x \in \Roots(\Gamma)} h \Downarrow s(x).
\]
Hence
\[
h = \bigcup_{x \in RC(\Gamma)} h \Downarrow_{s(RC(\Gamma))} s(x).
\]

We can show
\[
h_1 \supseteq \bigcup_{x \in RC(\Gamma_1)} h \Downarrow_{s(RC(\Gamma))} s(x)
\]
as follows.
Assume $\not\supseteq$ in order to show contradiction.
Then there is $a \notin h_1$ and $a$ in the right-hand side.
Hence $a \in h_2$.
Hence there is $x \in RC(\Gamma_1)$ such that $a \in h \Downarrow_{RC(\Gamma)} s(x)$.
Hence there is the path $s(x) \toto a$.
By going from $s(x)$ to $a$,
take the first $h_2$-element to be $c$. 
Let the previous element be $b$. 
Then $s(x) \toto b \to c \toto a$ and
$s(x) \toto b$ is in $h_1$.
From $s,h_1 \models \Gamma_1$ and $s,h_2 \models \Gamma_2$,
since $b \to c$ is from one cone to another cone,
we have $y \in \FV(\Gamma)$ such that $c=s(y)$.
Since $c \in h_2$, we have $y \in RC(\Gamma_2)$,
which contradicts with the path $s(x) \toto b \to s(y) \toto a$ in $h \Downarrow_{s(RC(\Gamma))} s(x)$ since $y \in RC(\Gamma)$.

Similarly we have
\[
h_2 \supseteq \bigcup_{x \in RC(\Gamma_2)} h \Downarrow_{s(RC(\Gamma))} s(x).
\]
Since 
\[
h_1+h_2 = 
\bigcup_{x \in RC(\Gamma_1) \cup RC(\Gamma_2)} h \Downarrow_{s(RC(\Gamma))} s(x),
\]
we have \[
h_i=\bigcup_{x \in RC(\Gamma_i)} h \Downarrow_{s(RC(\Gamma))} s(x).
\]
$\Box$

\begin{Lemma}\label{lemma:split}
(1)
If $\Gamma$ is $\Gamma_1*\Gamma_2$, and $\Gamma'$ is $\Gamma_1'*\Gamma_2'$, and
\[
h_1+h_2=h_1'+h_2', \\
\]
and for $i=1,2$,
\[
s,h_i \models \Gamma_i \land (\FV(\Gamma)-(\Roots+\Cells)(\Gamma_i))\uparrow, \\
s',h_i' \models \Gamma_i' \land  (\FV(\Gamma')-(\Roots+\Cells)(\Gamma_i'))\uparrow, \\
\Roots(\Gamma_i') \subseteq (\Roots+\Cells)(\Gamma_i) \subseteq (\Roots+\Cells)(\Gamma_i'),
\]
then
$h_i=h_i'$ for $i=1,2$.

(2)
If
\[
s,h_i \models \Gamma_i \ (i=1,2), \\
s,h_1+h_2 \models \Phi_1 * \Phi_2,
\]
and
$(\Gamma_1,\Gamma_2)$ is group-disjoint,
$(\Phi_1,\Phi_2)$ is a group split by $(\Gamma_1,\Gamma_2)$,
then
\[
s,h_i \models \Phi_i \ (i=1,2).
\]
\end{Lemma}

{\em Proof.}
(1)
Let $V_i$ be $(\Roots+\Cells)(\Gamma_i)$ and
$V_i'$ be $(\Roots+\Cells)(\Gamma_i')$ for $i=1,2$.
Let $V$ be $V_1 \cup V_2$ and
$V'$ be $V_1' \cup V_2'$.
Let $h$ be $h_1+h_2$.

By Lemma \ref{lemma:cone} (2),
\[
h_i' = \bigcup_{x \in V_i'} h \Downarrow_{s'(V')} s'(x).
\]

We can show 

Claim 1: $x \in V_1'-V_1$ implies $s(x) \in h_1$

as follows.
Assume $x \in V_1'-V_1$.
Then $x \in \Cells(\Gamma_1')$.
Then $s(x) \notin s(V)$ since $s(x) \in s(V_2)$ implies $s(x) \in s(V_2')$, which contradicts with $x \in V_1'$.
Moreover there is $y \in \Roots(\Gamma_1')$ such that
$P(y) \land x \downarrow$ is in $\Gamma_1'$.
Hence there is a path $s(y) \toto s(x)$ in $h_1'$.
By going from $s(y)$ to $s(x)$,
take $s(z)$ to be the last $s(V)$-element.
Then $z \in V$ and $s(y) \toto s(z) \to a \toto s(x)$ and
$a \toto s(x)$ are not in $s(V)$.
Then $z \in V_1$ since $z \notin V_1$ implies $z \in V_2$ and
$z \in V_2'$ from $V_2 \subseteq V_2'$, so $s(z) \in h_2'$,
which contradicts with $s(z) \in h_1'$.
Hence $s(x) \in h \Downarrow_{s(V)} s(z)$.
By Lemma \ref{lemma:cone} (3),
\[
h_1 = \bigcup_{x \in V_1} h \Downarrow_{s(V)} s(x).
\]
Hence $h_1 \ni s(x)$.
We have shown the claim 1.

We will show 

Claim 2: $h \Downarrow_{s(V')}s(x) \subseteq h_1$ for any $x \in V_1'$ 

as follows.
If $x \in V_1$, 
then $h \Downarrow_{s(V')}s(x) \subseteq h \Downarrow_{s(V)}s(x) \subseteq h_1$.
Assume $x \in V_1'-V_1$.
By the claim 1, $s(x) \in h_1$.
Hence there is $y \in V_1$ such that $s(x) \in h\Downarrow_{s(V)}s(y)$.
Hence $h\Downarrow_{s(V')}s(x) \subseteq h\Downarrow_{s(V)}s(x) \subseteq h\Downarrow_{s(V)}s(y) \subseteq h_1$.
We have shown the claim 2.

By the claim 2, $h_1 \supseteq h_1'$.
Similarly we have $h_2 \supseteq h_2'$.
Hence $h_i=h_i'$ for $i=1,2$.

(2)
Note that the roots are not bound in $\Phi_i$ for $i=1,2$
by the group split.

We have $h_1'+h_2' = h_1+h_2$ such that
\[
s,h_i' \models \Phi_i.
\]
Let $\Phi_i$ be $\exists\Vec x_i\exists\Vec y_i\Uparrow(\Pi_i\land \Gamma_i')$.
Then we have $\Vec a_i,\Vec b_i$ such that
by letting $s'$ be $s[\Vec x_i:=\Vec a_i,\Vec y_i:=\Vec b_i \ (i=1,2)]$,
we have
\[
s',h_i' \models \Gamma_i'.
\]
By (1), $h_i=h_i'$.
$\Box$

\section{Local Completeness of Rule $(\Bounded Factor)$ in Algorithm}

A rule is defined to be {\em locally complete} if
all its assumptions are valid when its conclusion is valid.

\begin{Lemma}\label{lemma:boundedfactorlocalcomplete}

Let $V,d$ be the arguments sent to {\rm MainLoop} from {\rm Main}.
Then, among processes produced by MainLoop, there is some process in which
every use of
the rule
\[
\infer[(\BFactor)]{x \ne y \land F \prove \Vec G,
G[(\Vec {Q(\Vec t)} \Swand P(x,\Vec t)) \land y\downarrow]}{
\hbox to 80ex{\vbox{\noindent $x \ne y \land F \prove \Vec G,
\{G[
\exists\Vec w(
(\Vec {Q_1(\Vec t_1)},Q(y,\Vec w) \Swand P(x,\Vec t)) *
(\Vec {Q_2(\Vec t_2)} \Swand Q(y,\Vec w))
)
]
\ |\
$
\hfil\break$
\{\Vec {Q_1(\Vec t_1)},\Vec {Q_2(\Vec t_2)}\} = \{\Vec Q(\Vec t)\},
Q \in \Dep(P), \Vec Q_2 \subseteq \Dep(Q), 
$\hfil\break$
|\Vec {Q_1(\Vec t_1)},Q(y,\Vec w)|_V \le d,|\Vec {Q_2(\Vec t_2)}|_V \le d,
\Vec w {\rm\ fresh}\}$
}}
}
\]
is locally complete.
\end{Lemma}

{\em Proof.}
Assume all the processes use some locally incomplete application of
$(\BFactor)$, in order for contradiction.

Consider the tree of processes where
each path represents a process,
the nodes are factor nodes and fork nodes,
which represent the application of the bounded factor rule
and fork with new processes respectively,
and any process is represented by some path.
We cut each path by the first application of locally incomplete
application of the bounded factor rule.
By the assumption, each path is cut.
By K\"onig's Lemma, there is the maximum depth of the cut tree.
Take a path of the maximum depth and consider the process represented by
this path.
Then for every choice for the application of the bounded factor rule by the process at the cut node is locally incomplete.
(Otherwise, by choosing some application that is locally complete,
the path can extend more than the maximum depth, which contradicts.)

For simplicity,
we write $v \Swand v'$ for $P'(v) \Swand P''(v')$ for some $P',P''$.
We also write $\{v_1,\ldots,v_n\} \Swand v$
for $P_1(v_1) \Swand \ldots \Swand P_n(v_n) \Swand P(v)$ for
some $P,P_1,\ldots,P_n$.
Note that any order of $\{v_1,\ldots,v_n\}$ gives the equivalent inductive
predicate by Lemma \ref{lemma:multiwand}.

Consider this process at the cut node.
We write $p$ for this process at the cut node.

Fix an application of the bounded factor rule in $p$.

Since the rule (\Factor) without the restriction on depth of wands
is locally complete by Lemma \ref{lemma:wand2},
we have some $\Vec\Phi$, $\Phi$, $\Vec\Phi'$, $\Phi''$,
such that
the entailment of the cut note is
the conclusion $\psi \prove \Vec\Phi,\Phi$ of the rule (\Factor)
and valid,
$\psi \prove \Vec\Phi,\Vec \Phi'$ is the assumption of the rule with
the restriction $d$ and invalid,
$\psi \prove \Vec\Phi,\Vec \Phi',\Vec\Phi''$ 
is the assumption of the rule without
the restriction and valid and contains some wand of depth $d+1$.
Then
there are $s',h'$ such that
\[
s',h' \models \psi, \\
s',h' \not\models \Vec\Phi,\Vec\Phi' \\
\]
Then there is $\Phi'' \in \Vec\Phi''$ such that
\[
s',h' \models \Phi''.
\]
Let 
\[
\Phi'' \equiv 
F[\exists\Vec w(S \Swand x) * y)]
\]
where $S$ is a set of variables and $y \in S$ and $|S-V|=d+1$.

Define $s$ from $s'$ by adding assignments for existential variables in $\Phi''$.
Then we have 
some $h'' \subseteq h'$ such that
\[
s,h'' \models S \Swand x.
\]
We obtain some initial heap $h$ by going back along 
the computation from $h'$ to the beginning in reverse order.

We call a variable $w$ an {\em extra} in an entailment $J$ when
$w \in \Roots(\psi)-\bigcap_i(\Roots(\Phi_i))$
where $J$ be $\psi \prove (\Phi_i)_{i \in I}$.
Assume $w \notin V$.
The wand $P(w) \Swand \ldots$ appears in the computation only when $w$ is an extra at some step.
$w$ is an extra at the step only when at the previous steps
we unfold some predicate with some root in the antecedent and
unfold another predicate with the same root in the succedent,
and 
we match some $\exists z$ in the antecedent and the corresponding $\exists w$
in the succedent.

By this observation, we make a sequence as follows:
We start with some $y \in V$ for $\exists\Vec w(y \Swand \ldots)$.
Then we take the next $w \in \Vec w$ for $\exists\Vec w'(w \Swand \ldots)$.
Then we similarly take the next $w' \in \Vec w'$ for $\exists\Vec w''(w' \Swand \ldots)$.
We repeat this.
Finally we collect all of such $w$ to make a sequence starting from $y$.
Note that there is not any $y' \in V$ such that $y'=w$ since $w \ne \FV$ by the algorithm.
Then in $h$ for the succedent,
we have a sequence $y,z_{1},\ldots,z_{d+1}$ such that
\[
y \in V, \\
z_{i} \notin V, \\
y \toto\to_\Bk z_{1}, \\
z_{i} \toto_\Tr \to_\Bk z_{i+1} \ \ (1 \le i \le d), \\
z_{i}' \to_\Tr z_{i} \ (1 \le j \le d+1), \\
z_{d+1}' \toto_\Tr \ldots \toto_\Tr z_{1}' \toto_\Tr y.
\]
Here $z_{i}$'s are different.
$z_{i}'$'s may be the same.
$z_{i} \ne z_{i}'$ for all $i$.
$z_{i}'$ may be in $V$.

Since an application of the bounded factor rule in $p$
is arbitrary,
we have these for each application of the bounded factor rule in $p$.
We index these applications by $I$.
Then in $h$ for the succedent,
for all $i \in I$,
we have sequences $y_i,z_{i1},\ldots,z_{i(d+1)}$ such that
\[
y_i \in V, \\
z_{ij} \notin V, \\
y_i \toto\to_\Bk z_{i1}, \\
z_{ij} \toto_\Tr \to_\Bk z_{i(j+1)} \ \ (1 \le j \le d), \\
z_{ij}' \to_\Tr z_{ij} \ (1 \le j \le d+1), \\
z_{i(d+1)}' \toto_\Tr \ldots \toto_\Tr z_{i1}' \toto_\Tr y_i.
\]
$z_{ij}'$'s may be the same.
$z_{ij} \ne z_{ij}'$ for all $i,j$.
$z_{ij}'$ may be in $V$.

Claim 1. $z_{ij}$'s are different.

It is because:
It is clear that $z_{ij}$'s are different for a fixed $i$.
For $i \ne k$,
$z_{ij}$ and $z_{kl}$ are different since
they belong to different heaps after some split step.

Claim 2.
In $h$ for the antecedent, $z_{ij} \ (1 \le j \le d)$ is not below $z_{k(d+1)}$.

It is because $z_{ij}$ is unfolded before $z_{k(d+1)}$ is unfolded.

Claim 3. The path
$
z_{i(d+1)}' \toto_\Tr \ldots \toto_\Tr z_{i1}' \toto_\Tr y_i
$
does not contain any $z_{kj}$.

We can show it as follows.
Assume the path contains some $z_{kj}$ in order for contradiction.
Consider some process $q$ such that
$q$ unfolds the same variables as $p$,
and $q$ unfolds $z_{kj}$ before $q$ does
the application of bounded factor indexed by $i$.
Then this application by $q$ is locally complete with depth $d$,
so the path of $q$ is longer than that of $p$, which contradicts with
the maximum length for $p$.
Hence we have shown the claim 3.

Claim 4. In $h$ for the antecedent,
every $z_{ij}'$ is below some $z_{k(d+1)}$.

We can show it as follows.
Assume some $z_{ij}'$ is not below $z_{k(d+1)}$ for all $k \in I$,
in order for contradiction.
Then $z_{ij}'$ is above or equals every $z$
such that the bounded factor for $z$ is locally incomplete.
Consider some process $q$ such that
$q$ unfolds the same variables as $p$,
and $q$ unfolds $z_{ij}'$ before $q$ does
the application of bounded factor indexed by $i$.
Then this application by $q$ is locally complete with depth $d$,
so the path of $q$ is longer than that of $p$, which contradicts with
the maximum length for $p$.
Hence we have shown the claim 4.

By the claim 1, the number of $z_{ij}$'s is $|I|(d+1)$.
Since the number of $z_{i(d+1)}$'s is $|I|$,
by the claims 2 and 4,
there is some $k \in I$ such that
$|S| \ge d+1$ where
$S$ is the set of $z_{ij}$ such that $z_{ij}'$ below $z_{k(d+1)}$
in $h$ for the antecedent.

By the claims 2 and 4,
the paths $z_i' \to z_i$ \ $(1 \le i \le k+1)$ are back edges
in $h$ for the antecedent.
Hence we have every element of $S$ in the arguments of the predicate for $z_{k(d+1)}$.
But $|S| \ge d+1 = k_{\max}+1$ and
the number of the arguments at $z_{k(d+1)}$ is $\le k_{\max}$, which contradicts.
$\Box$

\section{Local Completeness of Step Unrelatedness Introduction}

\begin{Lemma}\label{lemma:unrelated-aux}
If
\[
Y \uparrow \land \Pi \land \Gamma * 
\Delta_1 * \Delta_2 \models
\Vec\Phi,
\exists\Vec w(Y'\Uparrow \land \Pi' \land \Gamma' *
(\Gamma_1 * (\Delta \land w\downarrow)) * \Gamma_2), \\
w \in \Vec w, \\
(\Roots+\Cells)(\Gamma)=(\Roots+\Cells)(\Gamma')-\Vec w, \\
(\Roots+\Cells)(\Delta_1)=(\Roots+\Cells)(\Gamma_1 * (\Delta \land w\downarrow))-\Vec w, \\
(\Roots+\Cells)(\Delta_2)=(\Roots+\Cells)(\Gamma_2)-\Vec w,
\]
and 
$\exists\Vec w(Y'\Uparrow \land \Pi' \land \Gamma' *
(\Gamma_1 * (\Delta \land w\downarrow)) * \Gamma_2)$ is equality-full,
then
\[
Y \uparrow \land \Pi \land \Gamma * 
\Delta_1 * \Delta_2 \models
\Vec\Phi,
\exists\Vec w(Y'\Uparrow \land \Pi' \land \Gamma' * 
(\Gamma_1 * (\Delta \land w\downarrow))
* (\Gamma_2 \land w\Uparrow)).
\]
\end{Lemma}

{\em Proof.}
Assume $s,h$ satisfies the antecedent. Then we have $h_0+h_1+h_2=h$ such that
\[
s,h_0 \models \Gamma, \\
s,h_1 \models \Delta_1, \\
s,h_2 \models \Delta_2.
\]
If $s,h \models \Vec\Phi$, the claim immediately holds.
Assume 
\[
s,h \models
\exists\Vec w(Y'\Uparrow \land \Pi' \land \Gamma' *
(\Gamma_1 * (\Delta \land w\downarrow)) * \Gamma_2).
\]
Then we have $\Vec a,h_0'+h_1'+h_2'=h$ such that
by letting $s'$ be $s[\Vec w:=\Vec a]$,
\[
s',h_0' \models \Gamma', \\
s',h_1' \models \Gamma_1 * (\Delta \land w\downarrow), \\
s',h_2' \models \Gamma_2.
\]
By Lemma \ref{lemma:split} (1),
\[
h_0 = h_0', \\
h_1 = h_1', \\
h_2 = h_2'.
\]
Hence we have $w\uparrow$ for $h_2'$.
By Lemma \ref{lemma:leaves} and the equality-fullness, $s'(w) \notin \Leaves(h_2)$.
Hence 
\[
s',h_2' \models w\Uparrow.
\]
$\Box$

\begin{Lemma}\label{lemma:unrelatedness}
At the Unrelatedness Introduction step,
the rule
\[
\infer[({\rm UnrelatednessIntroduction})]{\psi \prove \Vec\Phi,
\exists\Vec x\exists\Vec y\Uparrow(\Pi \land \Gamma * (\Delta \land x\downarrow) * *_i\Gamma_i)}{
\psi \prove \Vec\Phi,
\exists\Vec x\exists\Vec y\Uparrow(\Pi \land \Gamma *(\Delta \land x\downarrow) * *_i(\Gamma_i \land x\Uparrow))
}
\\
\]
is locally complete.
\end{Lemma}

{\em Proof.}
Since $\Pi \land \Gamma * (\Delta \land x\downarrow) * *_i\Gamma_i)$
is grouping and equality-full at the step,
there are $\Delta_1,\Delta_2$ such that
the conditions of Lemma \ref{lemma:unrelated-aux} hold.
By Lemma \ref{lemma:unrelated-aux}, the claim holds.
$\Box$

\section{Selective Local Completeness of Rule $(*)$}

A set of rules of the same conclusion
is defined to be {\em selectively locally complete} if
for every valid conclusion of the rules
there is a locally complete rule in the set.

\begin{Prop}\label{prop:*localcomplete}
Let $S$ be the set of
\[
\infer[(\Split)]{Y \uparrow \land \Pi \land \Gamma_1 * \Gamma_2 \prove \{ \Phi_{1i} * \Phi_{2i} | i \in I \} }{
(Y \cup Y_1) \uparrow \land \Pi \land \Gamma_2 \prove \{ \Phi_{2i} | i \in I-I' \}
{\rm\ or\ }
	(Y \cup Y_2) \uparrow \land \Pi \land \Gamma_1 \prove \{ \Phi_{1i} | i \in I' \} 
	&
	(\forall I' \subseteq I)
}
\]
where
\[
V =\FV(Y \uparrow \land \Pi \land \Gamma_1 * \Gamma_2 \prove \{ \Phi_{1i} * \Phi_{2i} | i \in I \}), \\
Y_i = (\Roots+\Cells)(\Gamma_i)  \ (i=1,2), \\
\Pi \equiv (\ne (V)), \\
(\Roots+\Cells)(\Gamma_1*\Gamma_2) + Y = V, 
\]
$(\Gamma_1,\Gamma_2)$ is group-disjoint,
and
$(\Phi_{1i},\Phi_{2i}) \ (i \in I))$ is 
group split by $(\Gamma_1, \Gamma_2)$.
Then
$S$ is selectively locally complete,
namely,
if $Y \uparrow \land \Pi \land \Gamma_1 * \Gamma_2 \prove \{ \Phi_{1i} * \Phi_{2i} | i \in I \}$ is valid,
then
there is some rule in $S$ such that
all assumptions of the rule are valid.
\end{Prop}

{\em Proof.}
Assume
\[
Y \uparrow \land \Pi \land \Gamma_1 * \Gamma_2 \models \{ \Phi_{1i} * \Phi_{2i} | i \in I \}.
\]

Let
\[
\psi_1 = (Y \cup Y_2) \uparrow \land \Pi \land \Gamma_1, \\
\psi_2 = (Y \cup Y_1) \uparrow \land \Pi \land \Gamma_2.
\]

Fix $s,h_1,h_2$ and
assume 
\[
s,h_1 \models \psi_1, \\
s,h_2 \models \psi_2.
\]
Let $h=h_1+h_2$.
Then
\[
s,h \models Y \uparrow \land \Pi \land \Gamma_1 * \Gamma_2.
\]

Then we have
\[
\Lor_{i \in I} s,h_1+h_2 \models \Phi_{1i} * \Phi_{2i}.
\]

By Lemma \ref{lemma:split} (2),
\[
\Lor_{i \in I} (s,h_1 \models \Phi_{1i} \land s,h_2 \models \Phi_{2i}).
\]
By Lemma \ref{lemma:distribute}
\[
\Land_{I=I_1+I_2} ((\Lor_{i \in I_1} s,h_1 \models \Phi_{1i}) \lor
(\Lor_{i \in I_2} s,h_2 \models \Phi_{2i})).
\]
Since $s,h_1,h_2$ are arbitrary,
\[
\forall s h_1 h_2(s,h_1 \models \psi_1 \land s,h_2 \models \psi_2 \imp \\ \qquad
\Land_{I=I_1+I_2} ((\Lor_{i \in I_1} s,h_1 \models \Phi_{1i}) \lor
(\Lor_{i \in I_2} s,h_2 \models \Phi_{2i}))).
\]
Fix $I_1+I_2=I$. Then
\[
\forall s h_1 h_2(s,h_1 \models \psi_1 \land s,h_2 \models \psi_2 \imp \\ \qquad
((\Lor_{i \in I_1} s,h_1 \models \Phi_{1i}) \lor
(\Lor_{i \in I_2} s,h_2 \models \Phi_{2i}))).
\]
Hence
\[
\forall s h_1 h_2(
(s,h_1 \models \psi_1 \land s,h_2 \models \psi_2 \imp (\Lor_{i \in I_1} s,h_1 \models \Phi_{1i})) \lor \\ \qquad
(s,h_1 \models \psi_1 \land s,h_2 \models \psi_2 \imp (\Lor_{i \in I_2} s,h_2 \models \Phi_{2i}))).
\]

If $s,h_1 \models \psi_1 \land s,h_2 \models \psi_2$,
then 
$
\Lor_{i \in I_1} (s,h_1 \models \Phi_{1i}) \lor 
\Lor_{i \in I_2} (s,h_2 \models \Phi_{2i})
$.
If $s,h_1 \not\models \psi_1$, then
$
s,h_1 \models \psi_1 \imp \Lor_{i \in I_1} s,h_1 \models \Phi_{1i}
$.
If $s,h_2 \not\models \psi_1$, then
$
s,h_2 \models \psi_2 \imp \Lor_{i \in I_2} s,h_2 \models \Phi_{2i}
$.
Hence
\[
\forall s h_1 h_2(
(s,h_1 \models \psi_1 \imp \Lor_{i \in I_1} s,h_1 \models \Phi_{1i}) \lor
\\ \qquad
(s,h_2 \models \psi_2 \imp \Lor_{i \in I_2} s,h_2 \models \Phi_{2i})).
\]
Hence
\[
\forall s(\psi_1 \models_s \{\Phi_{1i} | i \in I_1\} \lor
\psi_2 \models_s \{\Phi_{2i} | i \in I_2\}).
\]
By Lemma \ref{lemma:constantstore}
\[
\psi_1 \models \{\Phi_{1i} | i \in I_1\} \lor
\psi_2 \models \{\Phi_{2i} | i \in I_2\}.
\]
Since $I_1+I_2=I$ are arbitrary, we have 
\[
\psi_1 \models \{\Phi_{1i} | i \in I_1\} \lor
\psi_2 \models \{\Phi_{2i} | i \in I_2\}
\]
for all $I_1+I_2=I$.

Consider the rule in $S$ such that
for each $I'$, by taking $I_1$ to be $I'$ and $I_2$ to be $I-I'$,
the assumption for $I'$ is taken to be the first disjunct when
\[
\psi_2 \models \{\Phi_{2i} | i \in I_2\}
\]
and is taken to be the second disjunct when
\[
\psi_1 \models \{\Phi_{1i} | i \in I_1\}.
\]
Then all the assumptions of this rule are valid.
$\Box$

\section{Fresh Variable in Succedent}

\begin{Lemma}\label{lemma:disjunct}
If $\theta$ is a variable renaming,
$
(\Dom \cup \Range)(\theta) \subseteq X$, 
and
$(\Dom \cup \Range)(\theta) \cap \FV(\Gamma) = \emptyset$, 
and
$\Pi \supseteq (\ne(\FV(X \uparrow \land \Pi \land \Gamma \prove \Vec \Phi,\Phi,\Phi\theta)))$
and $\Phi$ is equality-full,
then the rule
{\small
\[
\infer[({\rm FreshVariableDisjunctElim})]{
X \uparrow \land \Pi \land \Gamma \prove \Vec \Phi,\Phi,\Phi\theta}{
X \uparrow \land \Pi \land \Gamma \prove \Vec \Phi,\Phi
}
\]
}
is locally complete.
\end{Lemma}

{\em Proof.}
Let $\Vec x_1$ be $\FV(\Phi) \cap \Dom(\theta)$ and $\Vec x_2$ be $\theta(\Vec x_1)$.

Assume the conclusion of the rule is valid
in order to show the assumption of the rule is valid.
Assume $s,h \models X \uparrow \land \Pi \land \Gamma$
in order to show $\Vec\Phi,\Phi$.
Then $s,h \models \Vec \Phi,\Phi,\Phi\theta$.
Assume $s,h \models \Phi\theta$.
We will show $s,h \models \Phi$.

By Lemma \ref{lemma:leaves},
$\Leaves(h) \subseteq s(\FV(\Gamma) \cup\{\Nil\})$.
Hence
$s(\Vec x_1\Vec x_2) \notin \Leaves(h)$,
since $\Vec x_1\Vec x_2 \notin \FV(\Gamma)$.
Hence
$s(\Vec x_1\Vec x_2) \notin (\Dom \cup \Range)(h)$,
since $\Vec x_1\Vec x_2 \subseteq X$.

Let $T$ be $\FV(\Phi_1) \cup\{\Nil\}-\Vec x_1$.
Then
\[
s,h \models \exists \Vec x_2\Uparrow(\Phi\theta \land (\ne(\Vec x_2,T \cup \Vec x_2))).
\]
By variable renaming,
\[
s,h \models \exists \Vec x_1\Uparrow(\Phi \land (\ne(\Vec x_1,T \cup \Vec x_1)).
\]
Here $\exists \Vec x_1\Uparrow(\Phi \land (\ne(\Vec x_1,T \cup \Vec x_1))$
is equality-full.
By Lemma \ref{lemma:existsplit} (2),
\[
s[\Vec x_1:=s(\Vec x_1)],h \models \Phi \land (\ne(\Vec x_1,T \cup \Vec x_1)).
\]
Hence
\[
s,h \models \Phi.
\]
$\Box$

\section{Completeness of $\CSLIDomega$}

This section shows the completeness of $\CSLIDomega$
by using the algorithm, the local completeness, and the termination.

By the lemmas in previous sections,
we can show some process does not fail for a valid input.

\begin{Lemma}\label{lemma:nonfail}
(1) 
In the algorithm,
each step except the rules $(\BFactor)$ and $(\Split)$ in 
each function except Main and MainLoop 
transforms a valid entailment into
subgoals consisting of valid entailments.

(2) 
If a valid entailment is given to the algorithm,
for each call of the function MainLoop, there is some process such that
all applications of the rules $(\BFactor)$ and $(\Split)$ are
locally complete.

(3) If a valid entailment is given to the algorithm,
for each call of the function MainLoop, some process does not fail.
\end{Lemma}

{\em Proof.}
(1) We only discuss interesting cases.

- The function Split.

Step 4. Existential Disequality. 
We have $\ne$ for every variable except $\Vec z$.
$\Vec x \ne \Vec z$ from the antecedent.
$\Vec y \ne \Vec z$ from $\Vec y\uparrow$ and $\Vec z\downarrow$.

Step 5. Unrelatedness Introduction.
By Lemma \ref{lemma:unrelatedness}.

- Normalize

Step 1. Fresh Variable Disjunct Elimination.
By Lemma \ref{lemma:disjunct}. 

Step 2. Unnecessary Disequality Elimination.
Assume the antecedent of the assumption is true at $(s,h)$
in order to show the succedent of the assumption is true at $(s,h)$.
Choose $a \in \Val - \Dom(h)$ 
such that $a \notin s(\FV(\Pi,\Gamma,\Vec\Phi)\cup\{\Nil\})$.
Let $s'$ be $s[x:=a]$.
Then the antecedent of the conclusion is true at $(s',h)$.
Then the succedent of the conclusion is true at $(s',h)$.
Then the succedent of the assumption is true at $(s',h)$.
Since $y$ does not appear in the succedent,
the succedent of the assumption is true at $(s,h)$.

Step 3. Unnecessary Variable Elimination.
It is similar to Unnecessary Disequality Elimination
by choosing $a \notin \Dom(h)$.

Step 4. Unnecessary Unrelated Existential Elimination.
By Lemma \ref{lemma:existsUparrow}.

(2) By Proposition \ref{prop:*localcomplete} and Lemma \ref{lemma:boundedfactorlocalcomplete}.

(3) 
From (2), we have some process such that
all applications of the rules $(\BFactor)$ and $(\Split)$ are
locally complete.
Since Main does case analysis
and MainLoop does case analysis and discharges by bud-companion relation.
By (1), in this process,
applications of each rule are locally complete.
Hence all the entailments handled by this process are valid.
We show this process does not fail.
Assume it fails in order to show contradiction.
When the process fails, there is either some Termination Check step
in the function {\rm MainLoop} or
the step for checking $\Emp$ in the function {\rm Main}.
In both cases we have contradiction since
the entailment is shown to be invalid as follows:
For the function Main.
Since $\Pi \imp \Lor \Pi'$ is not true,
we have $s$ that satisfies $\Pi$ and $\Land \neg\Pi'$.
Take $h$ to be the empty heap.
$(s,h)$ is a counterexample since 
for every disjunct either $\Emp$ is not in it or
$\Pi'$ in it is false.
For MainLoop.
Since $\psi$ is satisfiable by the Unsat check step,
we have $s$ that satisfies $\Pi$.
Take $h$ to be the empty heap.
Then $(s,h)$ is a counterexample since $\Emp$ is not in the succedent.
$\Box$

Finally we can prove the completeness of $\CSLIDomega$.

\begin{Th}[Completeness]\label{th:complete}
(1) The system $\CSLIDomega$ is complete.
Namely,
if a given entailment $J$ is valid,
then it is provable in $\CSLIDomega$.

(2) The algorithm decides the validity of a given entailment.
Namely,
For a given input $J$,
the algorithm returns Yes when the input is valid,
and 
it returns No when the input is invalid.
\end{Th}

{\em Proof.}
(1) Assume $J$ is valid in order to show $J$ is provable in $\CSLIDomega$.
When we input $J$ to the algorithm,
by Lemma \ref{lemma:nonfail} (3), 
for each call of MainLoop, some process does not fail.
By Lemma \ref{lemma:terminate} (2), 
for each call of MainLoop, the  process terminates without fail.
Hence the algorithm terminates with Yes.
By Lemma \ref{lemma:partialcorrect}, $J$ is provable.

(2) Assume $J$ is valid, in order to show 
the algorithm with input $J$ terminates with Yes.
By Lemma \ref{lemma:nonfail} (3), 
for each call of MainLoop, some process does not fail.
By Lemma \ref{lemma:terminate} (2), the algorithm terminates with Yes.

Assume $J$ is invalid, in order to show 
the algorithm with input $J$ terminates with No.
By Lemma \ref{lemma:terminate} (2), the algorithm terminates.
Assume the algorithm with input $J$
terminates with Yes, in order to show contradiction.
By Lemma \ref{lemma:partialcorrect}, $J$ is provable.
By Theorem \ref{th:sound}, $\models J$, which contradicts.
Hence the algorithm with input $J$ terminates with No.
$\Box$

\section{Conclusion}

We have proposed
the cyclic proof system $\CSLIDomega$
for symbolic heaps with inductive definitions,
and have proved
its soundness theorem and
its completeness theorem, and
have given the decision procedure for the validity of a given entailment.

Future work would be to apply ideas in this paper to other systems,
in particular, the strong wand and 
the selective local completeness of the rule $(*)$.

\end{document}